\documentclass[preprint2]{aastex}

\newcommand{\bm}[1]{\mbox{\boldmath $#1$}}

\begin{document}

\title{Binary Neutron Stars in Quasi-equilibrium}

\author{Keisuke Taniguchi\altaffilmark{1}}
\affil{Department of Physics, University of Wisconsin-Milwaukee,
P.O. Box 413, Milwaukee, WI 53201, USA}
\altaffiltext{1}{Current address: Department of Earth Science and
Astronomy, Graduate School of Arts and Sciences,
University of Tokyo, Komaba, Meguro, Tokyo 153-8902, Japan}

\and

\author{Masaru Shibata}
\affil{Yukawa Institute for Theoretical Physics, Kyoto University,
Kyoto, 606-8502, Japan}

\begin{abstract}

Quasi-equilibrium sequences of binary neutron stars are constructed for
a variety of equations of state in general relativity. Einstein's
constraint equations in the Isenberg-Wilson-Mathews approximation are
solved together with the relativistic equations of hydrostationary
equilibrium under the assumption of irrotational flow. We focus on
unequal-mass sequences as well as equal-mass sequences, and compare
those results. We investigate the behavior of the binding energy and
total angular momentum along a quasi-equilibrium sequence, the endpoint
of sequences, and the orbital angular velocity as a function of time,
changing the mass ratio, the total mass of the binary system, and the
equation of state of a neutron star. It is found that the orbital
angular velocity at the mass-shedding limit can be determined by an
empirical formula derived from an analytic estimation. We also provide
tables for 160 sequences which will be useful as a guideline of
numerical simulations for the inspiral and merger performed in the
near future.

\end{abstract}


\section{Introduction}

Coalescing binary neutron stars are among the most promising sources
of gravitational waves for ground-based laser-interferometric
gravitational-wave detectors such as LIGO \citep{bro04}, GEO600
\citep{luc06}, TAMA300 \citep{and05}, and VIRGO \citep{ace07}. Merger
of binary neutron stars, together with that of black hole-neutron star
binaries, is also considered to be one of the candidates for the
central engines of short-hard gamma-ray bursts \citep{nar92}. These
facts motivate us to study coalescing binary neutron stars.

Binary neutron stars evolve as a result of gravitational radiation
reaction and eventually merge. This evolutionary sequence is usually
divided into three stages, depending on the characteristic timescales
associated with orbital period and gravitational radiation reaction, as
well as on the tidal effects for each neutron star. The first stage is
the adiabatic, inspiral phase. In this phase, the timescale of orbital
shrink due to the emission of gravitational waves is much longer than
the orbital period, and thus, the binary system evolves
adiabatically. In addition, each neutron star can be treated as a
point mass, because the orbital separation is much larger than the
neutron star radius and hence the tidal deformation of the neutron
star is negligible.  In this phase, a post-Newtonian approximation
together with the point particle approximation is a robust tool for
determining the orbital evolution and for computing gravitational
waveforms (see e.g. \citet{bla06} and references therein).

The second stage is called the intermediate phase or the
quasi-equilibrium phase. In this phase, the binary system is considered
to be still in the adiabatic, inspiral phase, but we need to take into
account tidal effects on each neutron star, i.e., hydrodynamic effects
in neutron stars, as well as full effects of general relativity,
because the orbital separation between two neutron stars is only a few
times of the neutron-star radius and thus they are in a strong
two-body gravitational field. One of the aims of the present paper is
to contribute to the understanding of this phase. We will explain more
details about the purpose of the present paper later.

The last stage is the merger phase, for which the timescale of orbital
shrink becomes shorter than the orbital period and thus the evolution
of the system proceeds in a dynamical manner. Furthermore, the system
becomes highly general relativistic, because the compactness of the
system, defined by the ratio of the gravitational radius to the radius
of the system, becomes larger than $\sim 0.2$. To clarify the merger
phase, numerical relativity is the unique approach. Since the first
fully general relativistic merger simulation was performed by
\citet{shi99}, huge effort has been devoted in this research field
\citep{shi00,shi02,shi03,shi05,shi06a,due03,mil04,and08a,and08b,yam08,liu08,bai08,gia09,kiu09,kiu10}.
(See e.g. \citet{oec07a} and \citet{oec07b} for works focusing on the
micro-physics in a neutron star but not in fully general relativistic
framework.)

Now we return to the intermediate phase and explain the purposes for
studying the quasi-equilibrium phase of binary neutron stars in general
relativity in detail. There are two roles for this study. One is to
clarify the physical conditions in this phase, for example, how large
the tidal deformation of a neutron star is, when the mass-shedding
from the neutron star occurs, and what the orbital angular velocity at
the mass-shedding limit is. The other is to provide initial data for
studying the merger phase by numerical relativity simulations.
Numerical relativity, in which Einstein's evolution equations are
solved, requires initial data that satisfy Einstein's constraint
equations and also that should be as physical as possible. Obviously,
it is important to derive accurate initial data for a scientific
study. Constructing such initial data is just obtaining relativistic
binary neutron stars in quasi-equilibrium.

The first effort on this issue was devoted to constructing corotating
binary neutron stars in general relativity because implementing a
numerical code for computing such solutions is relatively easy
\citep{bau97,bau98,mar98,usu00,usu02,tan02b,tan03,tic09}.
\citet{koc92} and \citet{bil92}, however, found that the timescale of
coalescence driven by the gravitational-radiation reaction is much
shorter than that of synchronization due to the viscosity in a neutron
star. This implies that if the spin angular velocity of neutron stars
is much smaller than the orbital angular velocity in a late inspiral
phase, we can regard the rotation state of a neutron star to be
approximately irrotational for the subsequent phase until the merger
sets in. Additionally, any neutron star spins down due to
electromagnetic radiation during its life from birth to the
coalescence. The spin down timescale of a neutron star in a known
binary is at longest as short as the coalescing timescale
\citep[larger than $\sim 10^8$ yrs;][]{lor08}.
Moreover, the spin period of neutron
stars in a known binary is always larger than 20 ms which is $\sim 10$
times larger than the orbital period in the late inspiral phase just
prior to merger, 2--3 ms. Therefore, we can conclude that the
irrotational flow is physically more realistic.\footnote[2]{If a first
  born neutron star in a binary was strongly
  recycled during the evolution of the companion, it may result in
  fast rotation and weak magnetic field. In such a case, the neutron
  star may rotate on the order of milliseconds even just before the
  merger. Some effort to approximately construct quasi-equilibrium,
  non-corotating, and non-irrotational binary systems is reported in
  \citet{mar03} and \citet{bau09}.}

Under the assumption of irrotation, formulation for solving
relativistic hydrostatic equations was derived
\citep{bon97,asa98,shi98,teu98}. Soon after the formulation was
derived, quasi-equilibrium sequences of irrotational binary neutron
stars were calculated
\citep{bon99,mar99,ury00a,ury00b,gou01,tan02b,tan03,bej05}, based on
the Isenberg-Wilson-Mathews (IWM) approximation to general relativity
\citep{ise78,ise08,wil89}. (See \citet{shi04} for an advanced
formulation and \citet{ury06,ury09} for the results.) Even though a
lot of sequences have been, so far, calculated, systematic survey has
not yet been done. In particular, {\it unequal-mass}, irrotational binary
neutron stars with an {\it equation of state other than single polytrope}
has not been studied in detail. In \citet{tan02b,tan03},
quasi-equilibrium sequences of unequal-mass binaries were calculated,
but a polytropic equation of state (EOS) was used. In \citet{bej05}
and \citet{ury09}, non-polytropic EOSs were used, but quasi-equilibrium
sequences of non-equal-mass binaries were not computed. Actually,
unequal-mass, irrotational binary neutron stars with realistic EOSs in
quasi-circular orbits were constructed and used initial data for
merger simulations in \citet{shi05}, \citet{shi06a}, and
\citet{kiu09,kiu10}.\footnote[3]{We use the term ``realistic equations
  of state'' for the EOSs derived from nuclear physics, although no
  one really knows a realistic one.} We, however, have not constructed
sequences, and rather computed only some initial data sets for each
neutron star mass. The purpose of the present paper is to complete the
issue and to provide a database of the sequences.

To compute unequal-mass binary systems of arbitrary mass ratio, we
develop a new code for the present research, because the numerical
code that we developed for the previous works \citep{tan02b,tan03}
had a problem with calculating binary systems composed of significantly
different-mass neutron stars in general relativity even though the
problem was not in Newtonian computation \citep{tan02a}. As we will 
explain in Section 2, the method to determine the center of mass of
unequal-mass binary systems, i.e., the position of the rotation axis,
caused the problem in the previous code, but we have overcome it by
employing a new method. 

In addition, we employ a wide variety of EOSs; piecewise polytropic
EOSs \citep{rea09a,rea09b}, tabulated realistic EOSs derived from
nuclear physics, and fitted EOSs to the tabulated realistic EOSs. Some
of the first and second EOSs were, respectively, used in \citet{ury09}
and in \citet{bej05}, but we adopt a wider set of EOSs in this paper.
Furthermore, we systematically study the unequal-mass binaries, whereas 
\citet{bej05} and \citet{ury09} focused only on the equal-mass case. 

This paper is organized as follows. We briefly review the basic
equations and explain the improvement of the numerical code in Section 2.
In Section 3, the results for the code test are shown. In Section 4,
we show numerical results and discuss the effects of EOS on each
sequence. Section 5 is devoted to a summary. Throughout this paper we
adopt geometrized units with $G=c=1$, where $G$ denotes the
gravitational constant and $c$ the speed of light. Latin and Greek
indices denote purely spatial and spacetime components, respectively.

\section{Formulation}

In this section, we briefly describe the basic equations to be solved
and the method to calculate a quasi-equilibrium configuration in circular
orbits. For more details, we would like to recommend readers to refer
to \citet{coo00}, \citet{bau03}, and \citet{gou07} for the part of
gravitational field equations, and \citet{gou01} for that of
hydrostatic equations.

\subsection{Gravitational field equations}

The line element in 3+1 form is written as
\begin{eqnarray}
  ds^2 &=& g_{\mu \nu} dx^{\mu} dx^{\nu}, \nonumber \\
  &=& -\alpha^2 dt^2 + \gamma_{ij} (dx^i +\beta^i dt)
  (dx^j +\beta^j dt), \nonumber \\
\end{eqnarray}
where $g_{\mu \nu}$ is the spacetime metric, $\alpha$ is the lapse
function, $\beta^i$ is the shift vector, and $\gamma_{ij}$ is the spatial
metric induced on a spatial hypersurface $\Sigma_t$. Note here that
the direction of the shift vector $\beta^i$ is the normally used one
which has a sign opposite to that used in \citet{gou01} and
\citet{tan02b,tan03}. The spatial metric $\gamma_{ij}$ is further
decomposed into the conformal factor $\psi$ and a background metric
$\tilde{\gamma}_{ij}$, and is written as
\begin{equation}
  \gamma_{ij} =\psi^4 \tilde{\gamma}_{ij}.
\end{equation}

The extrinsic curvature is defined by 
\begin{equation}
  K_{ij} =-\frac{1}{2} {\cal L}_n \gamma_{ij} \label{eq:extr_curv}
\end{equation}
where ${\cal L}_n$ is the Lie derivative along the unit normal to the
hypersurface $\Sigma_t$. We split it into the trace $K$ and the
traceless part $A_{ij}$ as
\begin{equation}
  K_{ij} =A_{ij} +\frac{1}{3} \gamma_{ij} K. \label{eq:extr_decomp}
\end{equation}
We further rewrite the traceless part as
\begin{mathletters}
\begin{eqnarray}
  A_{ij} &=& \psi^{-2} \tilde{A}_{ij}, \\
  A^{ij} &=& \psi^{-10} \tilde{A}^{ij}, \label{eq:extr_traceless}
\end{eqnarray}
\end{mathletters}
where $\tilde A_{ij} =\tilde{\gamma}_{ik} \tilde{\gamma}_{jl}
\tilde{A}^{kl}$.  The Hamiltonian constraint, then, takes the form
\begin{equation}
  \tilde{\nabla}^2 \psi =-2 \pi \psi^5 \rho_H
  +\frac{1}{8} \psi \tilde{R} +\frac{1}{12} \psi^5 K^2
  -\frac{1}{8} \psi^{-7} \tilde{A}_{ij} \tilde{A}^{ij},
  \label{eq:ham_constr}
\end{equation}
where $\tilde{\nabla}^2$ denotes
$\tilde{\gamma}^{ij} \tilde{\nabla}_i \tilde{\nabla}_j$,
$\tilde{\nabla}_i$ the covariant derivative with respect to
$\tilde{\gamma}_{ij}$, and $\tilde{R}$ the scalar curvature with
respect to $\tilde{\gamma}_{ij}$. The matter term, $\rho_H$, in
Equation (\ref{eq:ham_constr}) is calculated by taking a projection
of the stress-energy tensor. (See Equation (\ref{eq:rhoh}) below.)

Equations (\ref{eq:extr_curv}), (\ref{eq:extr_decomp}), and
(\ref{eq:extr_traceless}) yield
\begin{equation}
  \tilde{A}^{ij} =\frac{\psi^6}{2\alpha} \Bigl( \partial_t
  \tilde{\gamma}^{ij} +\tilde{\nabla}^i \beta^j
  +\tilde{\nabla}^j \beta^i -\frac{2}{3} \tilde{\gamma}^{ij}
  \tilde{\nabla}_k \beta^k \Bigr). \label{eq:traceless_extr}
\end{equation}
Using Equation (\ref{eq:traceless_extr}) for the traceless part of
the extrinsic curvature, the momentum constraint is written as
\begin{eqnarray}
  &&\tilde{\gamma}^{jk} \tilde{\nabla}_j \tilde{\nabla}_k \beta^i
  +\frac{1}{3} \tilde{\gamma}^{ik} \tilde{\nabla}_k
  (\tilde{\nabla}_j \beta^j) \nonumber \\
  &=&
  -\frac{\alpha}{\psi^6} \tilde{\nabla}_j \Bigl( \frac{\psi^6}{\alpha}
  \partial_t \tilde{\gamma}^{ij} \Bigr) +16 \pi \alpha \psi^4 j^i
  +\frac{4}{3} \alpha \tilde{\nabla}^i K \nonumber \\
  &&-\Bigl( \tilde{\nabla}^i \beta^j +\tilde{\nabla}^j \beta^i
  -\frac{2}{3} \tilde{\gamma}^{ij} \tilde{\nabla}_k \beta^k \Bigr)
  \frac{\alpha}{\psi^6} \tilde{\nabla}_j \Bigl( \frac{\psi^6}{\alpha}
  \Bigr), \nonumber \\
  \label{eq:mom_constr}
\end{eqnarray}
where the matter term, $j^i$, is calculated by taking a projection
of the stress-energy tensor. (See Equation (\ref{eq:ji}) below.)
In addition to the Hamiltonian and momentum constraints, we solve the
trace of the evolution equation of the extrinsic curvature
\begin{eqnarray}
  &&\partial_t K -{\cal L}_{\beta} K = -\psi^{-4}
  (\tilde{\nabla}_i \tilde{\nabla}^i \alpha + 2 \tilde{\nabla}_i
  \ln \psi \tilde{\nabla}^i \alpha) \nonumber \\
  &&\hspace{15pt} + \alpha \Bigl[ 4 \pi (\rho_H +S) +\psi^{-12}
    \tilde{A}_{ij} \tilde{A}^{ij} +\frac{K^2}{3} \Bigr],
  \label{eq:trace_evol}
\end{eqnarray}
where $S$ is a matter term defined by Equation (\ref{eq:s}) below.

As mentioned, the matter terms in the above equations are calculated
by taking the projections of the stress-energy tensor $T_{\mu \nu}$
into the spatial hypersurface $\Sigma_t$. In the present paper,
we assume an ideal fluid and adopt the form of $T_{\mu \nu}$ as
\begin{equation}
  T_{\mu \nu} =(\rho +\rho \epsilon +P) u_{\mu} u_{\nu}
  +P g_{\mu \nu},
\end{equation}
where $u_{\mu}$ is the fluid 4-velocity, $\rho$ is the baryon rest-mass
density, $\epsilon$ is the specific internal energy, and $P$ is the
pressure. Defining the future-directed unit normal to $\Sigma_t$ as
$n_{\mu}$, the projections of $T_{\mu \nu}$ can be written as
\begin{eqnarray}
  \rho_H &=& n_{\mu} n_{\nu} T^{\mu \nu}, \label{eq:rhoh} \\
  j^i &=&- \gamma_{\mu}^i n_{\nu} T^{\mu \nu}, \label{eq:ji} \\
  S_{ij} &=&\gamma_{i \mu} \gamma_{j \nu} T^{\mu \nu}, \\
  S &=& \gamma^{ij} S_{ij}. \label{eq:s}
\end{eqnarray}

The set of equations, (\ref{eq:ham_constr})--(\ref{eq:trace_evol}),
has four functions that we can choose freely;
$\partial_t \tilde{\gamma}^{ij}$, $\partial_t K$,
$\tilde{\gamma}_{ij}$, and $K$. For simplicity, we chose a maximal
slicing $K=0$ and adopt a flat metric
$\tilde{\gamma}_{ij}=\eta_{ij}$
for the spatial background metric in the present paper.
We then assume the presence of a helical
Killing vector, $\xi^{\mu}$, and the absence of gravitational waves in
the wavezone. Under these assumptions, it is natural to choose the
time direction so as to satisfy
$\xi^{\mu}=(\partial/\partial t)^{\mu}$, and to set
$\partial_t \tilde{\gamma}^{ij}=0$ and $\partial_t K=0$. Then, the 
basic equations are written as
\begin{eqnarray}
  &&\underline{\Delta} \psi =-2 \pi \psi^5 \rho_H
  -\frac{1}{8} \psi^{-7} \tilde{A}_{ij} \tilde{A}^{ij},
  \label{eq:psi} \\
  &&\underline{\Delta} \beta^i +\frac{1}{3} \eta^{ik} \partial_k
  (\partial_j \beta^j) =16 \pi \Phi \psi^3 j^i \nonumber \\
  &&\hspace{95pt} +2 \tilde{A}^{ij} \partial_j (\Phi \psi^{-7}),
  \label{eq:beta} \\
  &&\underline{\Delta} \Phi =2 \pi \Phi \psi^4 (\rho_H +2S)
  +\frac{7}{8} \Phi \psi^{-8} \tilde{A}_{ij} \tilde{A}^{ij},
  \label{eq:phi} \\
  &&\tilde{A}^{ij}=\frac{\psi^7}{2\Phi} \Bigl( \eta^{ik} \partial_k
  \beta^j +\eta^{jk} \partial_k \beta^i -\frac{2}{3} \eta^{ij}
  \partial_k \beta^k \Bigr), \nonumber \\
  \label{eq:taij}
\end{eqnarray}
where $\underline{\Delta}$ denotes the flat Laplacian, $\partial_i$
the flat partial derivative, and $\Phi \equiv \alpha \psi$. 
The derived formulation is called the IWM approximation
\citep{ise78,ise08,wil89}. Note here that a similar but more general
formulation was recently proposed and started to be called the
extended conformal thin-sandwich decomposition \citep[XCTS;][]{pfe03},
mainly in the field of vacuum spacetime, i.e., black hole
spacetime. In this new formulation, the conformal flatness is not
assumed. (See \citet{gou07} for a more detailed explanation.)

It may be worthy to note that there is an effort to construct
quasi-equilibrium binary neutron stars beyond the IWM
approximation. \citet{shi04} proposed a formalism in which all the
Einstein equations are solved under the assumption of a helical
Killing vector and artificially imposing asymptotic flatness. This
formalism was used to solve quasi-equilibrium binary neutron stars in
\citet{ury06,ury09}.

When solving the set of equations, (\ref{eq:psi})--(\ref{eq:taij}),
we need to impose the outer
boundary conditions for $\psi$, $\beta^i$, and $\Phi$. Because our
numerical grids include spatial infinity by compactifying the
outermost domain, we can impose the exact outer boundary conditions as
\begin{eqnarray}
  \psi &=& 1, \\
  \beta^i &=& ({\bm \Omega} \times {\bm R})^i, \\
  \Phi &=& 1,
\end{eqnarray}
where ${\bm \Omega}$ is the orbital angular velocity and ${\bm R}$ is
the radial coordinate measured from the center of mass of the binary
system. Although the shift vector seen by a co-orbiting observer,
$\beta^i$, diverges at infinity, this diverging term,
\begin{equation}
  \beta_{\rm rot}^i \equiv ({\bm \Omega} \times {\bm R})^i,
  \label{eq:shift_rot}
\end{equation}
does not affect the set of equations we solve. If we define the shift
vector seen by an inertial observer as $\beta_{\rm iner}^i$, the shift
vector seen by a co-orbiting observer can be written as
\begin{equation}
  \beta^i = \beta_{\rm iner}^i + \beta_{\rm rot}^i.
  \label{eq:shift_decomp}
\end{equation}
Because we have the relations,
\begin{eqnarray}
  &&\underline{\Delta} \beta_{\rm rot}^i = 0, \\
  &&\partial_j \beta_{\rm rot}^j = 0,
\end{eqnarray}
and
\begin{equation}
  \eta^{ik} \partial_k \beta_{\rm rot}^j +\eta^{jk} \partial_k
  \beta_{\rm rot}^i =0,
\end{equation}
substituting Equation (\ref{eq:shift_decomp}) into Equations
(\ref{eq:beta}) and (\ref{eq:taij}) yields
\begin{eqnarray}
  &&\underline{\Delta} \beta_{\rm iner}^i +\frac{1}{3} \eta^{ik}
  \partial_k (\partial_j \beta_{\rm iner}^j) =16 \pi \Phi \psi^3 j^i
  \nonumber \\
  &&\hspace{95pt} +2 \tilde{A}^{ij} \partial_j (\Phi \psi^{-7}), \\
  &&\tilde{A}^{ij}=\frac{\psi^7}{2\Phi} \Bigl( \eta^{ik} \partial_k
  \beta_{\rm iner}^j +\eta^{jk} \partial_k \beta_{\rm iner}^i
  \nonumber \\
  &&\hspace{50pt} -\frac{2}{3} \eta^{ij} \partial_k
  \beta_{\rm iner}^k \Bigr).
\end{eqnarray}
The outer boundary condition for $\beta_{\rm iner}^i$ is then
$\beta_{\rm iner}^i=0$ at spatial infinity. We actually solve for
$\beta_{\rm iner}^i$.

\subsection{Hydrostatic equations}

The hydrostatic equations governing the quasi-equilibrium state are the
Euler and continuity equations. For both irrotational and synchronized
motions, the Euler equation can be integrated once to give 
\begin{equation}
  h \alpha \frac{\gamma}{\gamma_0} ={\rm constant}, \label{eq:euler}
\end{equation}
where $h=(\rho +\rho \epsilon +P)/\rho$ is the fluid specific
enthalpy, $\gamma_0$ is the Lorentz factor between the co-orbiting and
Eulerian observers, and $\gamma$ is the Lorentz factor between the fluid
and co-orbiting observers. If we define the 4-velocity of the
co-orbiting observer by $v^{\mu}$, the Lorentz factors are written as
\begin{eqnarray}
  \gamma_0 &=& -n^{\mu} v_{\mu}
  = (1 -\gamma_{ij} U_0^i U_0^j)^{-1/2}, \\
  \gamma &=& -v^{\mu} u_{\mu} \nonumber \\
  &=& \gamma_0 (1 -\gamma_{ij} U^i U_0^j)
  (1 -\gamma_{ij} U^i U^j)^{-1/2},
\end{eqnarray}
where $U_0^i$ is the orbital 3-velocity with respect to the Eulerian
observer,
\begin{equation}
  U_0^i =\frac{\beta^i}{\alpha}, \label{eq:u0}
\end{equation}
and $U^i$ denotes the fluid 3-velocity with respect to the Eulerian
observer,
\begin{equation}
  U^i =\frac{\psi^{-4}}{\alpha u^t h} \tilde{\nabla}^i \Psi,
\end{equation}
for irrotational binary systems. Here $u^t$ is the time component of
the fluid 4-velocity and $\Psi$ is the velocity potential which is
calculated by solving the equation of continuity written as
\begin{equation}
  \frac{\rho}{h} \nabla^{\mu} \nabla_{\mu} \Psi
  +(\nabla^{\mu} \Psi) \nabla_{\mu} \Bigl( \frac{\rho}{h} \Bigr)
  =0,
\end{equation}
where $\nabla_{\mu}$ is the covariant derivative with respect to
$g_{\mu \nu}$. Note that the fluid 3-velocity $U^i$ corresponds to the
orbital 3-velocity $U_0^i$ for synchronized binary systems.

For the determination of the constant on the right-hand side of
Equation (\ref{eq:euler}), we use the central value of the quantities on
its left-hand side. The center of a neutron star is defined as the
location of the maximum baryon rest-mass density in the present
paper. Equation (\ref{eq:euler}) includes one more constant which
should be determined for each quasi-equilibrium figure; the constant is
the orbital angular velocity as we find from Equations (\ref{eq:u0}),
(\ref{eq:shift_decomp}), and (\ref{eq:shift_rot}). The method for 
calculating it will be explained in the next section.

\subsection{Orbital angular velocity and the center of mass of a
binary system}

The method for determining the orbital angular velocity is as follows:
we first set the rotation axis of the binary system to be the
$Z$-axis, and the line connecting the centers of mass of each neutron star
to be the $X$-axis. Requiring a force balance along the $X$-axis, we
impose a condition of quasi-circular orbit for the binary system. The
force balance equation is obtained by setting the central values of
the gradient of enthalpy to be zero for each star,
\begin{equation}
  \frac{\partial \ln h}{\partial X} \Bigl|_{{\cal O}_a} \Bigr.= 0,
  \label{eq:forcebalance}
\end{equation}
where ${\cal O}_a$ ($a=1,2$) denotes the center of each neutron
star. Because Equation (\ref{eq:euler}) includes $\Omega$ through
Equation (\ref{eq:shift_rot}), the force balance Equation
(\ref{eq:forcebalance}) may be regarded as the equation for
determining the orbital angular velocity. Equation
(\ref{eq:forcebalance}) also depends on the location of the center of
mass, because Equation (\ref{eq:shift_rot}) includes ${\bm R}$ which is the
radial coordinate measured from the center of mass of the binary
system. For equal-mass binaries, the force balance equations for each
star degenerate and the location of the center of mass becomes
trivial, i.e., we can set it to the middle between two stars. For
unequal-mass binaries, on the other hand, we have a couple of
equations, Equation (\ref{eq:forcebalance}) for $a=1$ and 2, for two
parameters of the orbital angular velocity and the location of the
center of mass. In the previous papers \citep{tan02b,tan03}, those
parameters were determined by solving the couple of equations as
stated in Section II B of \citet{tan02a}. This method works for Newtonian
binary systems \citep{tan02a} and also in the case that the difference
in mass of the neutron stars is small for general relativistic binary
systems. However, if the difference in mass of the neutron stars is
significantly large, the coupled equations, Equation (\ref{eq:forcebalance})
for $a=1$ and 2, would not be simultaneously satisfied at earlier steps of
computational iteration because the state of the binary neutron stars
is far from equilibrium, and as a result, the computation would fail
to achieve the convergence to a solution.

To avoid such crush of computation for small mass ratios,
$M_{\rm ADM}^{\rm NS 1}/M_{\rm ADM}^{\rm NS 2} < 0.8$ where
$M_{\rm ADM}^{\rm NS a}~(a=1,2)$ denotes the Arnowitt-Deser-Misner
(ADM) mass for a spherical star $a$ in isolation, we adopt the same
method as used for black hole-neutron star binaries, described in 
\citet{tan06,tan07,tan08}, to determine the location of the center of
mass; we require that the linear momentum of the system vanishes 
\begin{equation}
  P^i =\frac{1}{8 \pi} \oint_{\infty} K^{ij} dS_j=0. \label{eq:linemom}
\end{equation}
Here we have assumed maximal slicing condition, $K=0$. Once the
location of the center of mass is determined in an iteration step,
we move the position of each star, keeping the separation, in order
for the center of mass of the binary system to locate on the $Z$-axis.

For the computation of the orbital angular velocity, we keep the
method that Equation (\ref{eq:forcebalance}) is satisfied. As the readers
may realize, Equation (\ref{eq:forcebalance}) gives two values of the
orbital angular velocity for the unequal-mass case because there are
two equations in Equation (\ref{eq:forcebalance}). Even though those values
of the orbital angular velocity are very close (the relative
difference is within the convergence level), they are slightly
different because of numerical error. In the present paper, we just
take an average of the two values.

It may be worthy to comment on another method for determining the
orbital angular velocity. While our method for calculating $\Omega$ is
to require the force balance Equation (\ref{eq:forcebalance}), it is
possible to obtain $\Omega$ by requiring the enthalpy at two points on
the neutron star's surface to be equal, i.e., $h=1$ on the surface. We
confirm that the results of those two methods coincide within the
convergence level of the enthalpy.

\subsection{Global quantities and a mass-shedding indicator}

A sequence of binary neutron stars should be constructed for a 
fixed baryon rest mass of each star,
\begin{equation}
  M_{\rm B}^{(a)} =\int_{{\rm star}~a} \rho u^t \sqrt{-g} d^3 x,
  \hspace{10pt} a=1,2,
\end{equation}
as the orbital separation decreases. This is because we regard
the baryon rest mass as conserved as the orbital separation decreases
due to the emission of gravitational waves. Along such a
constant-baryon-rest-mass sequence, we then monitor three global
quantities: the ADM mass, the Komar mass, and the total angular
momentum, as well as a sensitive mass-shedding indicator of a star 
(see Equation (\ref{eq:defchi})).

The ADM mass in isotropic Cartesian coordinates is written as
\begin{equation}
  M_{\rm ADM} = -\frac{1}{2\pi} \oint_{\infty} \partial^i \psi dS_i.
  \label{eq:admsurf}
\end{equation}
If we use Equation (\ref{eq:psi}) and Gauss' theorem, the ADM mass can be
written in terms of volume integral as
\begin{equation}
  M_{\rm ADM} = \int_V \Bigl( \psi^5 \rho_H +\frac{1}{16\pi}
  \psi^{-7} \tilde{A}_{ij} \tilde{A}^{ij} \Bigr) dV. \label{eq:admvol}
\end{equation}
Both of Equations (\ref{eq:admsurf}) and (\ref{eq:admvol}) give the same
results relative to the convergence level of the computation. 

The Komar mass is written as
\begin{equation}
  M_{\rm Komar} = \frac{1}{4\pi} \oint_{\infty} \partial^i \alpha
  dS_i, \label{eq:Komarsurf}
\end{equation}
where we use the fact that the shift vector falls off rapidly enough
to be neglected from Equation (\ref{eq:Komarsurf}). Using the boundary
conditions that $\Phi=1$ and $\psi=1$ at infinity and the definition
$\Phi \equiv \alpha \psi$, we can rewrite Equation (\ref{eq:Komarsurf}) as
\begin{equation}
  M_{\rm Komar} = \frac{1}{4\pi} \oint_{\infty}
  (\partial^i \Phi - \partial^i \psi) dS_i.
\end{equation}
Using Equations (\ref{eq:psi}) and (\ref{eq:phi}), the Komar mass can be
also written in terms of volume integral as
\begin{eqnarray}
  M_{\rm Komar} &=& \frac{1}{4\pi} \int_V \Bigl[ 2\pi \psi^4 (\Phi+\psi)
    \rho_H +4\pi \Phi \psi^4 S \nonumber \\
   && +\frac{1}{8} \psi^{-7}
    (7 \Phi \psi^{-1} +1) \tilde{A}_{ij} \tilde{A}^{ij} \Bigr] dV.
\end{eqnarray}
The total angular momentum of the binary system is calculated to give 
\begin{equation}
  J_i =\frac{1}{16\pi} \epsilon_{ijk} \oint_{\infty}
  (X^j K^{kl} -X^k K^{jl}) dS_l,
\end{equation}
where $X^i$ is a spatial Cartesian coordinate relative to the center
of mass of the binary system. This equation can be also rewritten in
the form of volume integral as
\begin{equation}
  J_i = \epsilon_{ijk} \int_V \psi^{10} X^j j^k dV,
\end{equation}
where we use the momentum constraint equation.
Similarly, the linear momentum (\ref{eq:linemom}) is written as
\begin{equation}
  P^i = \int_V \psi^{10} j^i dV.
\end{equation}
As we mentioned, both of the global quantities calculated by surface
integral at infinity and by volume integral give the same results
within the convergence level of the computation. In the present paper,
we show the results by the volume integral.

The binding energy of the binary system is defined as
\begin{equation}
  E_{\rm b} =M_{\rm ADM} -M_0,
\end{equation}
where $M_0$ is the ADM mass of the binary system at infinite orbital
separation, as defined by the sum of the ADM mass of two isolated
neutron stars with the same baryon rest mass,
\begin{equation}
  M_0 \equiv M_{\rm ADM}^{\rm NS 1} +M_{\rm ADM}^{\rm NS 2}.
\end{equation}

To measure a global error in the numerical results, we define
the error in the virial theorem as the fractional difference between
the ADM and Komar masses,
\begin{equation}
  \delta M \equiv \Bigl| \frac{M_{\rm ADM} -M_{\rm Komar}}
  {M_{\rm ADM}} \Bigr|.
\end{equation}
We refer to $\delta M$ as the virial error, and use it to measure the
magnitude of numerical error in the ADM mass.

Finally, a sensitive mass-shedding indicator is defined as
\begin{equation}
  \chi \equiv \frac{(\partial (\ln h)/ \partial r)_{\rm eq}}
  {(\partial (\ln h)/ \partial r)_{\rm pole}}. \label{eq:defchi}
\end{equation}
Here, the numerator of Equation (\ref{eq:defchi}), $(\partial (\ln
h)/\partial r)_{\rm eq}$, is the radial derivative of the enthalpy in
equatorial plane at the surface along the direction toward the
companion star, and the denominator, $(\partial (\ln h)/\partial
r)_{\rm pole}$, is that at the surface of the pole. The radial
coordinate $r$ is measured from the center of the corresponding
neutron star. For spherical stars at infinite separation, the
indicator takes $\chi=1$, while $\chi=0$ indicates the formation of a
cusp, and hence the onset of mass-shedding. Note here that because our
numerical code is based on a spectral method, it is impossible to
construct cusp-like configurations, because it is accompanied by the
Gibbs phenomena.  This is also the case for the configuration with
smaller values of $\chi \leq 0.5$.  Thus, we stop constructing a
sequence when $\chi$ reaches $\sim 0.6$.

\vspace{0.8cm}
\begin{figure}[ht]
\epsscale{1.0}
\plotone{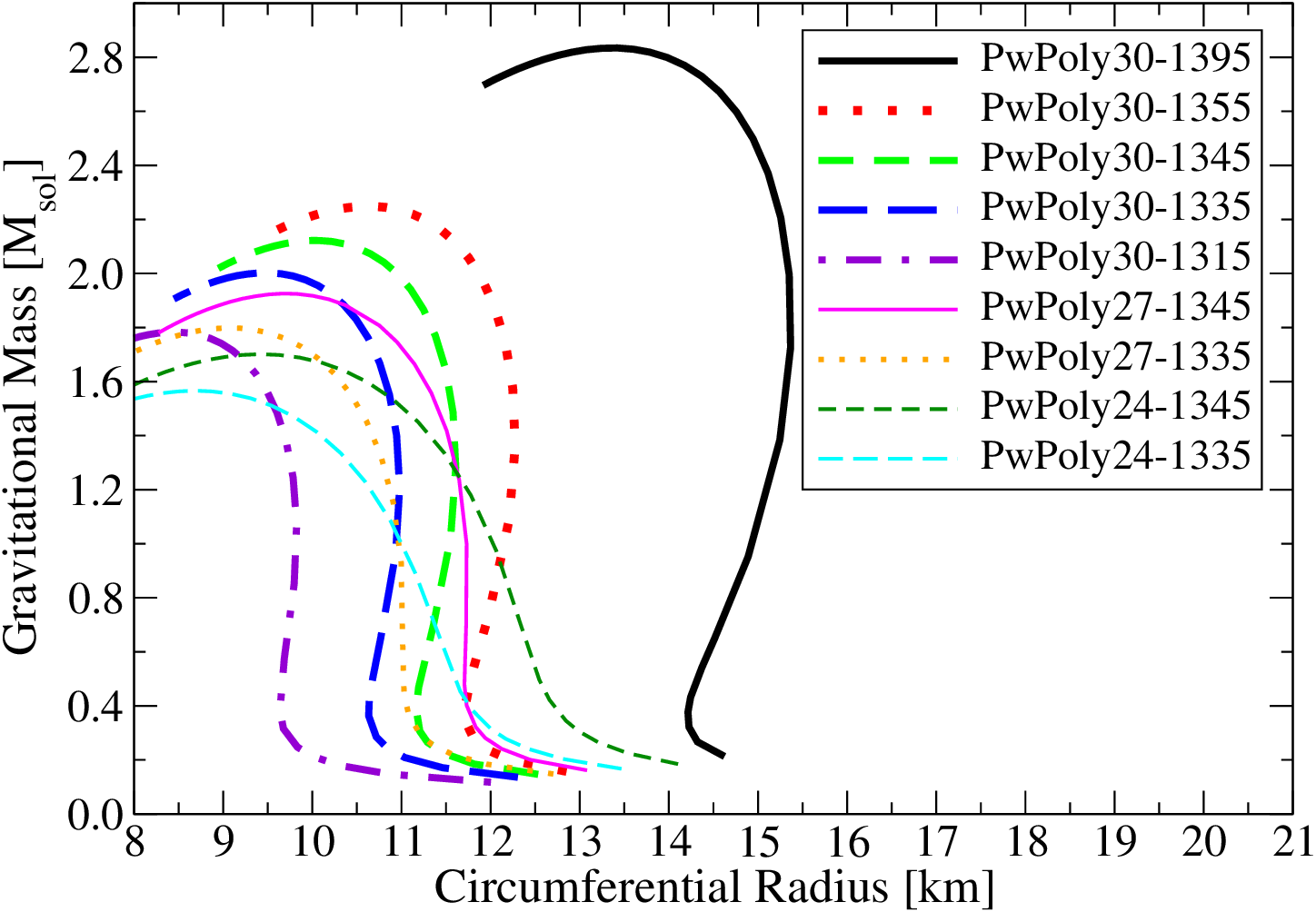}
\caption{Mass-radius relation for spherical stars with piecewise
  polytropic EOSs. The vertical axis is the gravitational mass (which
  is the same as the ADM mass) in solar mass units and the horizontal
  axis is the circumferential radius in km units. The thick (black)
  solid , thick (red) dotted, thick (green) short-dashed, thick (blue)
  long-dashed, and thick (violet) dot-dashed curves denote the case of
  $\Gamma_1=3.0$ but $\log_{10} P_1=13.95$ (PwPoly30-1395), 13.55
  (PwPoly30-1355), 13.45 (PwPoly30-1345), 13.35 (PwPoly30-1335), and
  13.15 (PwPoly30-1315), respectively. The thin (magenta) solid and
  thin (orange) dotted curves are the case of $\Gamma_1=2.7$ but
  $\log_{10} P_1=13.45$ (PwPoly27-1345) and 13.35 (PwPoly27-1335),
  respectively. The thin (dark green) short-dashed and thin (cyan)
  long-dashed curves denote the case of $\Gamma_1=2.4$ but
  $\log_{10} P_1=13.45$ (PwPoly24-1345) and 13.35 (PwPoly24-1335),
  respectively.
\label{fig1}}
\end{figure}

\subsection{Equations of state}

In this section, we summarize four types of EOSs employed in this 
paper.

\subsubsection{Polytropic EOSs}

The first EOS is a polytrope 
\begin{equation}
  P = \kappa \rho^{\Gamma},
\end{equation}
where $\kappa$ is a polytropic constant and $\Gamma$ is the adiabatic
index. This EOS has been often used in simply modeling binary 
neutron stars in quasi-equilibrium. We use this EOS only in testing 
our new code. The specific internal energy for the polytrope is
written as
\begin{equation}
  \epsilon =\frac{\kappa \rho^{\Gamma-1}}{\Gamma-1}.
\end{equation}

For the polytropic EOS, we have the following natural units, i.e.,
polytropic units, to normalize the length, mass, and time scales: 
\begin{equation}
  R_{\rm poly} =\kappa^{1/(2\Gamma -2)}.
\end{equation}
Because geometrized units with $G=c=1$ are adopted, the polytropic units
$R_{\rm poly}$ normalize all of the length, mass, and time scales.

\begin{deluxetable}{ccccc}
\tabletypesize{\normalsize}
\tablecaption{Parameters for the piecewise polytropic EOSs
\label{table1}}
\tablewidth{0pt}
\tablehead{
\colhead{Name} & \colhead{$\Gamma_1$} & \colhead{$\log_{10} P_1$} &
\colhead{$\rho_0~[{\rm g~cm^{-3}}]$} & \colhead{$a_1$}
}
\startdata
PwPoly30-1395 & 3.0 & 13.95 & $7.03317468 \times 10^{13}$ &
$8.06036645 \times 10^{-3}$ \\
PwPoly30-1355 & 3.0 & 13.55 & $1.23196176 \times 10^{14}$ &
$9.84569621 \times 10^{-3}$ \\
PwPoly30-1345 & 3.0 & 13.45 & $1.41728987 \times 10^{14}$ &
$1.03506910 \times 10^{-2}$ \\
PwPoly30-1335 & 3.0 & 13.35 & $1.63049750 \times 10^{14}$ &
$1.08815874 \times 10^{-2}$ \\
PwPoly30-1315 & 3.0 & 13.15 & $2.15795830 \times 10^{14}$ &
$1.20264673 \times 10^{-2}$ \\
PwPoly27-1345 & 2.7 & 13.45 & $1.06888797 \times 10^{14}$ &
$9.00037733 \times 10^{-3}$ \\
PwPoly27-1335 & 2.7 & 13.35 & $1.26878530 \times 10^{14}$ &
$9.56832301 \times 10^{-3}$ \\
PwPoly24-1345 & 2.4 & 13.45 & $6.85371121 \times 10^{13}$ &
$7.24283128 \times 10^{-3}$ \\
PwPoly24-1335 & 2.4 & 13.35 & $8.54665331 \times 10^{13}$ &
$7.83658359 \times 10^{-3}$ \\
\enddata
\end{deluxetable}

\subsubsection{Piecewise polytropic EOSs}

The second EOS is a piecewise polytrope introduced by
\citet{rea09a,rea09b}. In the present paper, we set the number of
polytrope segments to two. Then, the EOS is written as
\begin{eqnarray}
  P &=& \kappa_0 \rho^{\Gamma_0},
  \hspace{5pt} (0 \le \rho < \rho_0) \\
  P &=& \kappa_1 \rho^{\Gamma_1},
  \hspace{5pt} (\rho_0 \le \rho)
\end{eqnarray}
where the dividing density $\rho_0$ is close to the nuclear density of
order $\sim 10^{14}~{\rm g~cm^{-3}}$ (see below). The adiabatic index of
the crust is set to a fixed value as $\Gamma_0=1.35692895$, whereas we
choose three values for the adiabatic index of the core,
$\Gamma_1=2.4$, 2.7, and 3.0. The polytropic constant of the crust,
$\kappa_0$, is set to $\kappa_0/c^2=3.99873692 \times 10^{-8}
~(({\rm g~cm^{-3}})^{1-\Gamma_0})$ in cgs units. The polytropic
constant of the core, $\kappa_1$, is calculated by requiring the
continuity of the pressure at the dividing density $\rho_0$ as
\begin{equation}
  \kappa_1 =\kappa_0 \rho_0^{\Gamma_0 -\Gamma_1}. \label{eq:kappa1}
\end{equation}
The dividing density $\rho_0$ is calculated by setting the fiducial
density, $\rho_1$, and the pressure at the fiducial density, $P_1$. We
take the fiducial density as $\log_{10} \rho_1 =14.7$ where $\rho_1$
is in cgs units. Because $\rho_1$ is larger than the dividing density
$\rho_0$, the EOS has the form of $P_1 =\kappa_1 \rho_1^{\Gamma_1}$ at
the fiducial density. Using this equation and Equation (\ref{eq:kappa1}),
the dividing density is obtained as
\begin{equation}
  \log_{10} \rho_0 =
    \frac{\log_{10} P_1 -14.7 \times \Gamma_1 -\log_{10} \kappa_0}
    {\Gamma_0 -\Gamma_1}.
\end{equation}

The specific internal energy in the crust and that in the core
are, respectively, written as
\begin{eqnarray}
  \epsilon_0 &=& \frac{\kappa_0 \rho^{\Gamma_0-1}}{\Gamma_0-1} \\
  \epsilon_1 &=& a_1 +\frac{\kappa_1 \rho^{\Gamma_1-1}}{\Gamma_1-1},
\end{eqnarray}
where $a_1$ is a constant which is calculated by requiring the
continuity of the enthalpy at the dividing density $\rho_0$.

We summarize the adiabatic index of the core $\Gamma_1$, the logarithm
of the pressure at the fiducial density $\log_{10} P_1$, the dividing
density $\rho_0$, and the constant $a_1$ in Table \ref{table1}; we
employ nine parameter sets in the present work.  Figure \ref{fig1}
plots the mass-radius relation of spherical stars for the chosen
EOSs and indicates that a wide variety of EOSs are modeled.

\subsubsection{Tabulated realistic EOSs}

We use tabulated EOSs for zero-temperature nuclear matter which are
derived by using various theories of dense nuclear matter and
different solution methods of the many-body problem in nuclear
physics. As described in \citet{bej05}, the EOS of \citet{bay71} is
used for $\rho < 10^8~{\rm g~cm^{-3}}$, that of \citet{hae94}
for $10^8~{\rm g~cm^{-3}} < \rho < 10^{11}~{\rm g~cm^{-3}}$, and that
obtained by \citet{dou01} for
$10^{11}~{\rm g~cm^{-3}} < \rho < \rho_{\rm cc}$
in the neutron star crust, where
$\rho_{\rm cc} =(0.6 - 1.4) \times 10^{14}~{\rm g~cm^{-3}}$
is the density at the crust-core interface. (See \citet{cha08} for a
review of the EOS in the neutron star crust.)

\begin{figure}[ht]
\epsscale{1.0}
\plotone{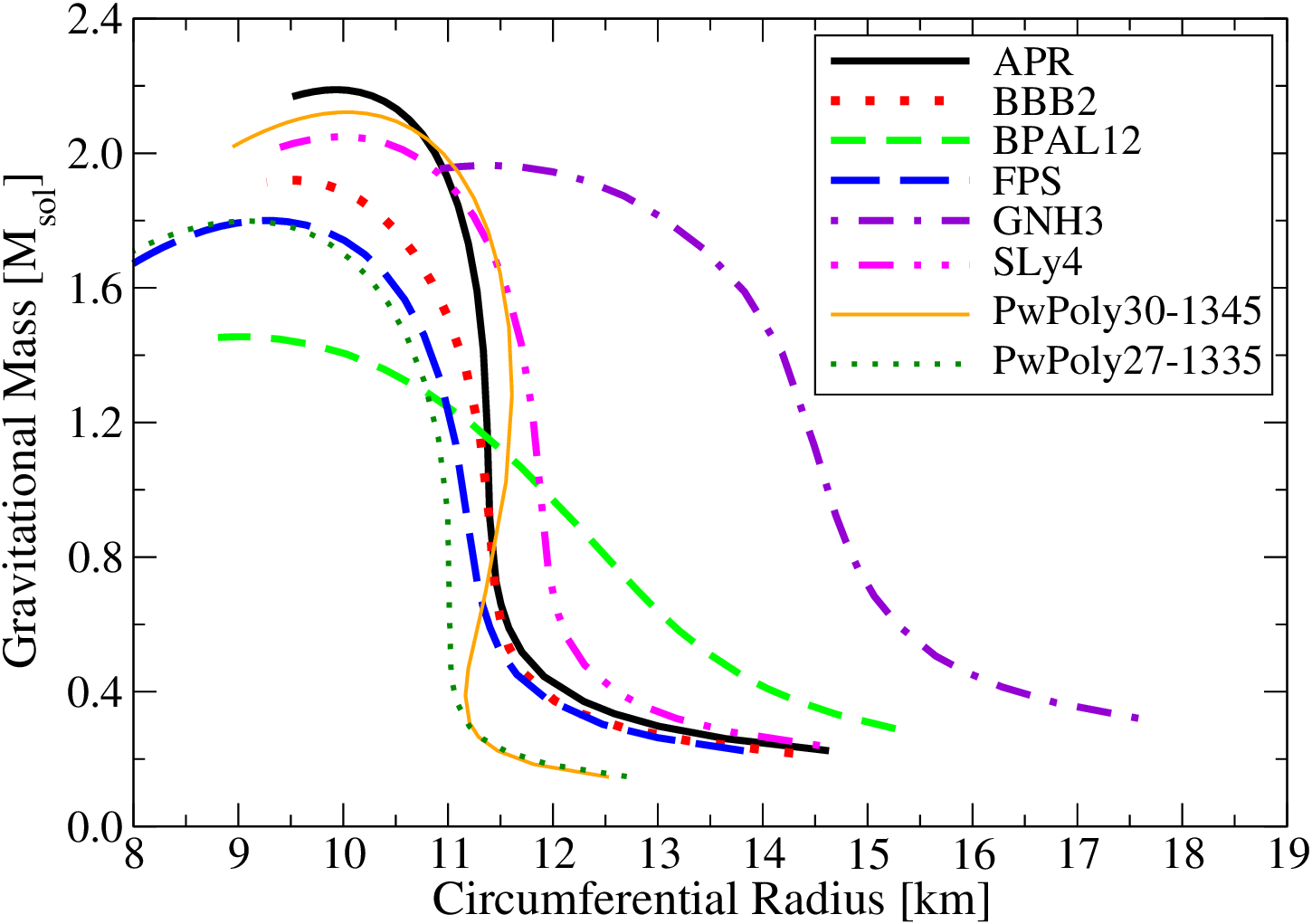}
\caption{Same as Figure \ref{fig1} but for tabulated realistic EOSs. The
  thick (black) solid, thick (red) dotted, thick (green) short-dashed,
  thick (blue) long-dashed, thick (violet) dot-dashed, and thick
  (magenta) dot-dot-dashed curves, respectively, denote the case of
  APR, BBB2, BPAL12, FPS, GNH3, and SLy4. The thin (orange) solid and
  thin (dark green) dotted curves are the case of piecewise polytropic
  EOSs with $\Gamma_1 =3.0$ and $\log_{10} P_1=13.45$ (PwPoly30-1345)
  and $\Gamma_1 =2.7$ and $\log_{10} P_1=13.35$ (PwPoly27-1335). Those
  curves are shown for comparison.
\label{fig2}}
\end{figure}

For the EOS of the neutron star core, we select six EOSs: APR
\citep{akm98}, BBB2 \citep{bal97}, BPAL12 \citep{zuo99}, FPS
\citep{fri81}, GNH3 \citep{gle85}, and SLy4 \citep{dou01}. Those
tabulated EOSs are interpolated by using the Hermite interpolation
which is basically the same method as in \citet{swe96}. 
Figure \ref{fig2} shows the mass-radius relation for spherical stars 
in these EOSs.

\subsubsection{Fitted EOSs to the tabulated realistic EOSs}

The method of an interpolation for the tabulated realistic EOSs is not
unique. \citet{hae05} introduced a fitting formula using an analytic
function. We modified their method \citep{shi05} and derived a new
fitting formula for FPS and SLy4 to satisfy the first law of
thermodynamics. In \citet{shi06a}, we also derived a fitting formula
for APR. In the present paper, we construct quasi-equilibrium sequences
for those EOSs, fitAPR, fitFPS, and fitSLy4, and compare the results
with those by the tabulated EOSs. 

\begin{figure}[ht]
\epsscale{1.0}
\plotone{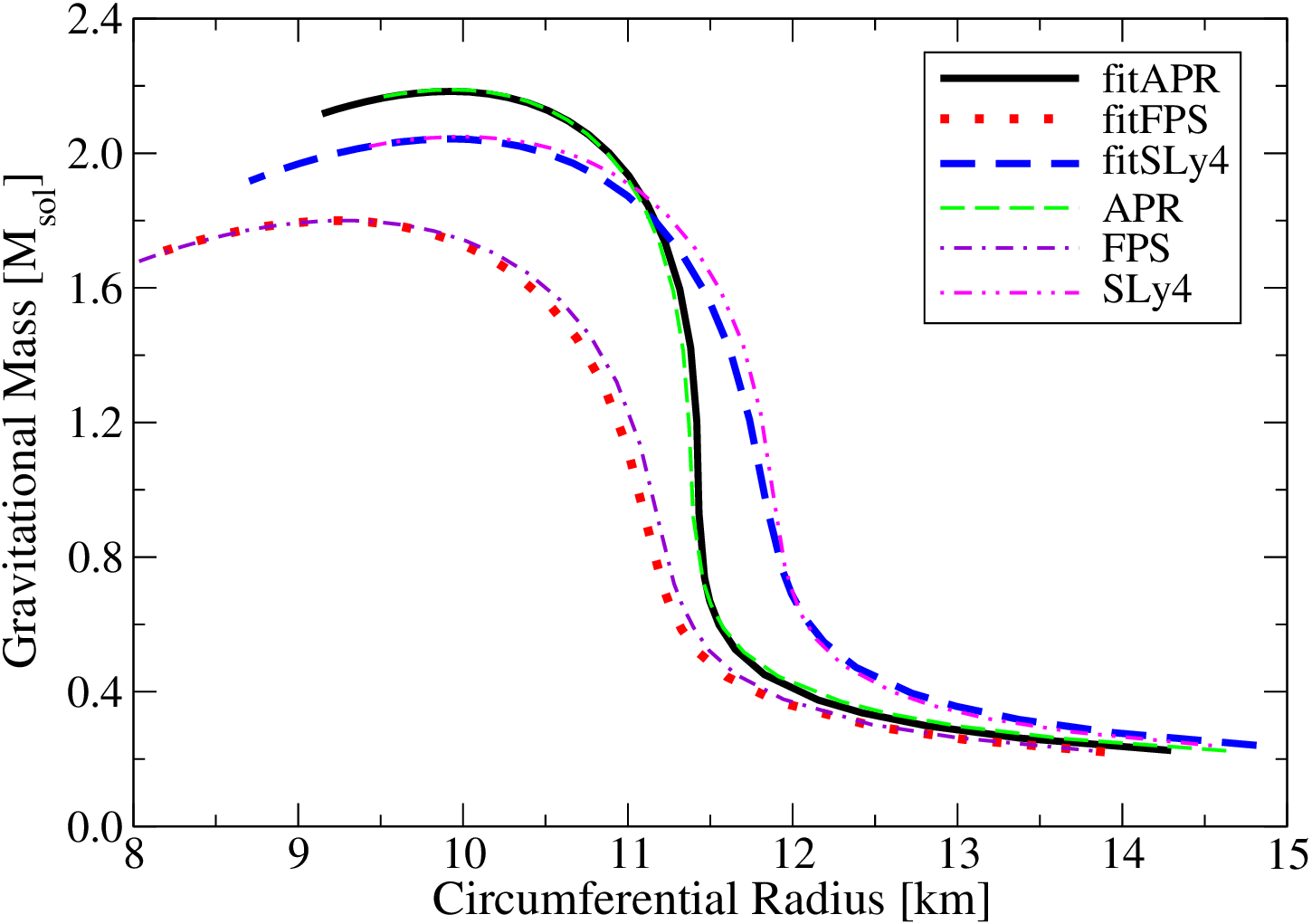}
\caption{Same as Figure \ref{fig1} but for EOSs written by a fitting
  formula. The thick (black) solid, thick (red) dotted, and thick
  (blue) short-dashed curves denote the case of fitAPR, fitFPS, and
  fitSLy4, respectively. The thin (green) long-dashed, thin (violet)
  dot-dashed, and thin (magenta) dot-dot-dashed curves are the case
  of tabulated EOSs, APR, FPS, and SLy4, respectively.
\label{fig3}}
\end{figure}

Figure \ref{fig3} compares the mass-radius relation for spherical
stars with those by the tabulated EOSs, APR, FPS, and SLy4.  A good
agreement is found between two corresponding results for each EOS. 

\section{Code Tests}

We implemented a new numerical code based on the spectral method
library, {\sc LORENE}, developed by the Meudon relativity group.  (See
the LORENE Web site, http://www.lorene.obspm.fr/, for more detailed
explanations of this code library.) Our new numerical code was tested
to check its ability for accurately computing unequal-mass binary
systems composed of significantly different-mass neutron stars as well
as equal-mass ones with similar accuracy to those obtained by the old
code used in \citet{gou01}, \citet{tan02b,tan03}, and \citet{bej05}.

\begin{figure}[ht]
\epsscale{1.0}
\plotone{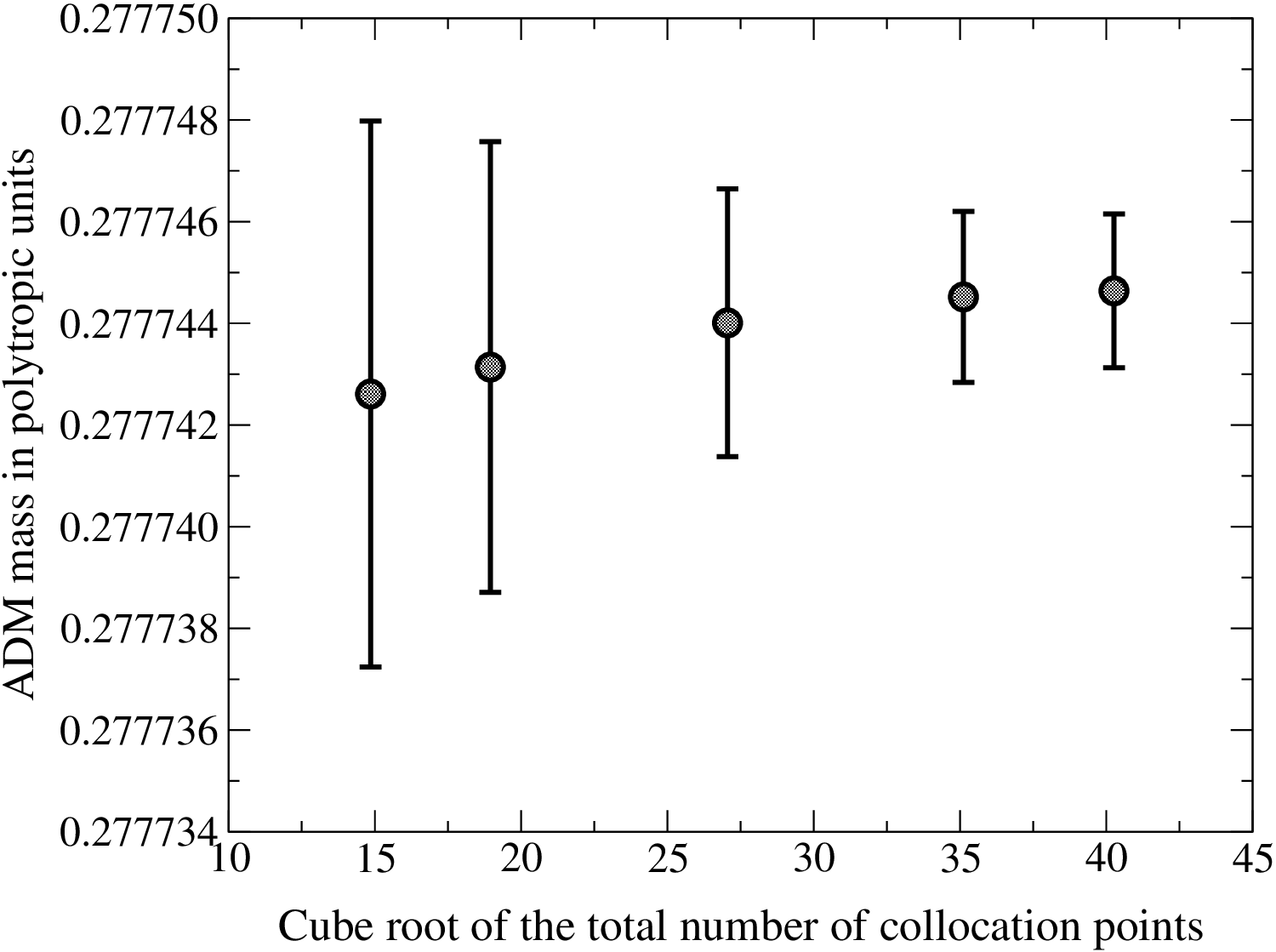}
\caption{Convergence test of the ADM mass of a binary neutron star
  with the $\Gamma=2$ polytropic EOS. The baryon rest mass of each
  star is $\bar{M}_{\rm B}=0.15$ and the coordinate orbital separation
  is set to $\bar{d}=7.308$ ($d/a_{\rm NS}=8.965$). The horizontal
  axis denotes the cube root of the total number of collocation
  points. The filled circles are the central value of each
  computation, and the error bars are drawn for an estimated error
  size derived from the virial error.
\label{fig4}}
\end{figure}

\begin{figure}[ht]
\epsscale{1.0}
\plotone{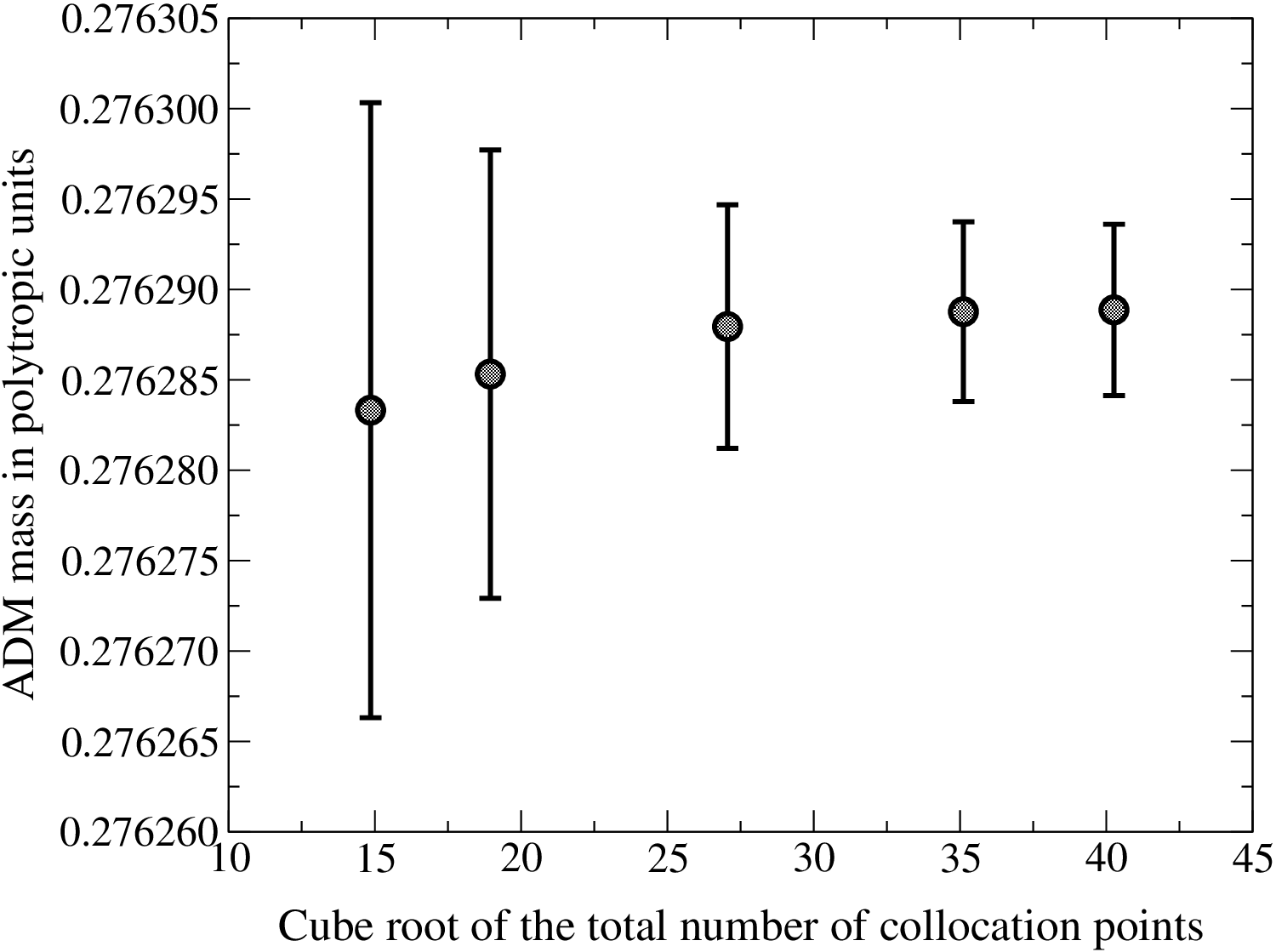}
\caption{Same as Figure \ref{fig4} but for the coordinate, orbital
  separation of $\bar{d}=3.045$ ($d/a_{\rm NS}=3.736$).
\label{fig5}}
\end{figure}

\subsection{Convergence tests}

The first test is to check if the convergence of the ADM mass of an
equal-mass binary neutron star is achieved with increasing the 
resolution. We choose the $\Gamma=2$ polytrope and set the baryon
rest mass of each star to $\bar{M}_{\rm B}=0.15$ in polytropic
units. Two different orbital separations, $\bar{d}=7.308$ and 3.045 in
polytropic units, are chosen. Those coordinate separations are,
respectively, about 8.965 and 3.736 times larger than the coordinate
radius of an isolated, spherical neutron star with the same baryon
rest mass. We choose these separations because the former one,
$d/a_{\rm NS}=8.965$, is larger than the farthest separation in the
equal-mass sequences we show in the present paper
($d/a_{\rm NS} \sim 8.6$) and the latter one, $d/a_{\rm NS}=3.736$,
is an intermediate separation between the farthest and the closest
ones.

Figures \ref{fig4} and \ref{fig5} compare the ADM mass at 
$\bar{d}=7.308$ and $3.045$, respectively, for different
resolutions (for different number of collocation points in the
terminology of the spectral method). We choose five resolutions,
$N_r \times N_{\theta} \times N_{\phi}=$ $49 \times 37 \times 36$,
$41 \times 33 \times 32$, $33 \times 25 \times 24$,
$25 \times 17 \times 16$, and $21 \times 13 \times 12$, where $N_r$,
$N_{\theta}$, and $N_{\phi}$ denote the number of collocation points
for the radial, polar, and azimuthal directions, respectively. The
horizontal axis of these figures denotes the cube root of the total
number of collocation points,
$\sqrt[3]{N_r \times N_{\theta} \times N_{\phi}}$. The error bar is
drawn for an estimated error size derived from the virial error.

It is found from Figure \ref{fig4} that the ADM mass for $41 \times 33
\times 32$ ($\sqrt[3]{N_r \times N_{\theta} \times N_{\phi}} \simeq
35.11$) is in approximately convergent level because its value is
approximately identical to that for $49 \times 37 \times 36$
($\sqrt[3]{N_r \times N_{\theta} \times N_{\phi}} \simeq 40.26$).  The
central value of the ADM mass for $33 \times 25 \times 24$ 
($\sqrt[3]{N_r \times N_{\theta} \times N_{\phi}} \simeq 27.05$) is in
the error bar of that for $49 \times 37 \times 36$. This implies that
the number of collocation points, $33 \times 25 \times 24$, is
sufficiently large when computing the binary with smaller separations
than $d/a_{\rm NS} \sim 9.0$ within the fractional error of $10^{-5}$
for the ADM mass, which satisfies the required accuracy in our present
computation.

Figure \ref{fig5} shows that the ADM mass for $33 \times 25 \times 24$
is in approximately convergent level because its value is
approximately identical to that for $49 \times 37 \times 36$ at the
separation of $\bar{d}=3.045$ ($d/a_{\rm NS}=3.736$). The central
value for $25 \times 17 \times 16$
($\sqrt[3]{N_r \times N_{\theta} \times N_{\phi}} \simeq 18.95$)
is in the error bar of that for $49 \times 37 \times 36$. This implies
that it is safe to decrease the number of collocation points for
smaller separations than $d/a_{\rm NS} \sim 3.7$ within the accuracy
required in the present computation.

From these convergence tests, we decide to use the number of
collocation points of $N_r \times N_{\theta} \times N_{\phi}=33 \times
25 \times 24$ for larger separations and $33 \times 17 \times 16$ for
closer ones, keeping the number of collocation points for the radial
direction.

\begin{figure}[ht]
\epsscale{1.0}
\plotone{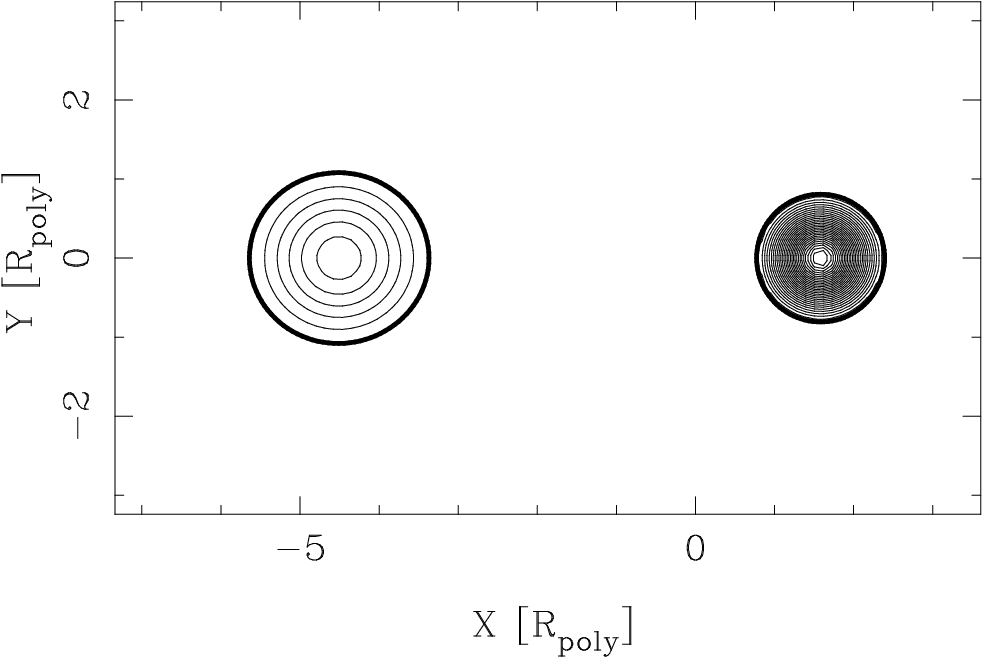}
\caption{Contours of baryon rest-mass density for demonstrating the
  ability of constructing a quasi-equilibrium figure composed of
  significantly different-mass stars. The EOS we employ is the
  $\Gamma=2$ polytrope. The star on the left-hand side has the baryon
  rest mass of $\bar{M}_{\rm B}=0.05$, while that on the right-hand
  side has $\bar{M}_{\rm B}=0.15$. The thick solid circles are the
  location of stellar surface.
\label{fig6}}
\end{figure}

\subsection{An unequal-mass binary system composed of significantly
different-mass stars}

To illustrate that our new code can compute a binary system composed
of significantly different-mass stars, we show in Figure \ref{fig6} a 
quasi-equilibrium configuration of binary neutron stars with the
$\Gamma=2$ polytrope whose baryon rest masses are
$\bar{M}_{\rm B}=0.05$ and 0.15, respectively. The compactness of
those stars when they have a spherical shape is, respectively,
${\cal C} \equiv M_{\rm ADM}^{\rm NS}/R_{\rm NS}=0.04155$ and
0.1452, where $R_{\rm NS}$ is the circumferential radius. Because the
maximum baryon rest mass of the $\Gamma=2$ polytrope is
$\bar{M}_{\rm B} \approx 0.18$, the model of $\bar{M}_{\rm B}=0.05$ is
an extremely light neutron star. We compute this model just to
demonstrate the ability of our code. In the figure, the star on the
left-hand side has $\bar{M}_{\rm B}=0.05$ and that on the right-hand
side does 0.15. The thick solid circles are the location of the
neutron star's surface. The orbital angular velocity, the binding
energy, the total angular momentum, the virial error, and the linear
momentum of this figure are $M_0 \Omega = 5.218 \times 10^{-3}$,
$E_{\rm b}/M_0 = -2.816 \times 10^{-3}$,
$J/M_0^2 = 1.1626$, $\delta M = 2.027 \times 10^{-6}$,
and $P^Y/(M_0 c) = 3.930 \times 10^{-7}$, respectively.
The relative difference of the binding energy from that obtained by
the third post-Newtonian (3PN) approximation at the same orbital
angular velocity is about
$\delta E_{\rm b} = -6.479 \times 10^{-4}$. Here we define
\begin{equation}
  \delta E_{\rm b} \equiv \frac{(E_{\rm b})_{\rm num}}
    {(E_{\rm b})_{\rm 3PN}} - 1, \label{eq:deb}
\end{equation}
where $(E_{\rm b})_{\rm num}$ and $(E_{\rm b})_{\rm 3PN}$ denote
the binding energy obtained by the numerical computation and that by
the 3PN approximation, respectively. The relative difference of the
total angular momentum from that obtained by the 3PN approximation at
the same orbital angular velocity is
$\delta J = -1.693 \times 10^{-4}$, defining $\delta J$ similar to
Equation (\ref{eq:deb}). The relative error in baryon rest mass of the less
massive star is $\delta \bar{M}_{\rm B}=6.738 \times 10^{-7}$ and that
of the more massive star is $2.082 \times 10^{-6}$. We do not find any
problem to accurately construct a binary system composed of two
significantly different-mass neutron stars.

\begin{figure}[ht]
\epsscale{1.0}
\plotone{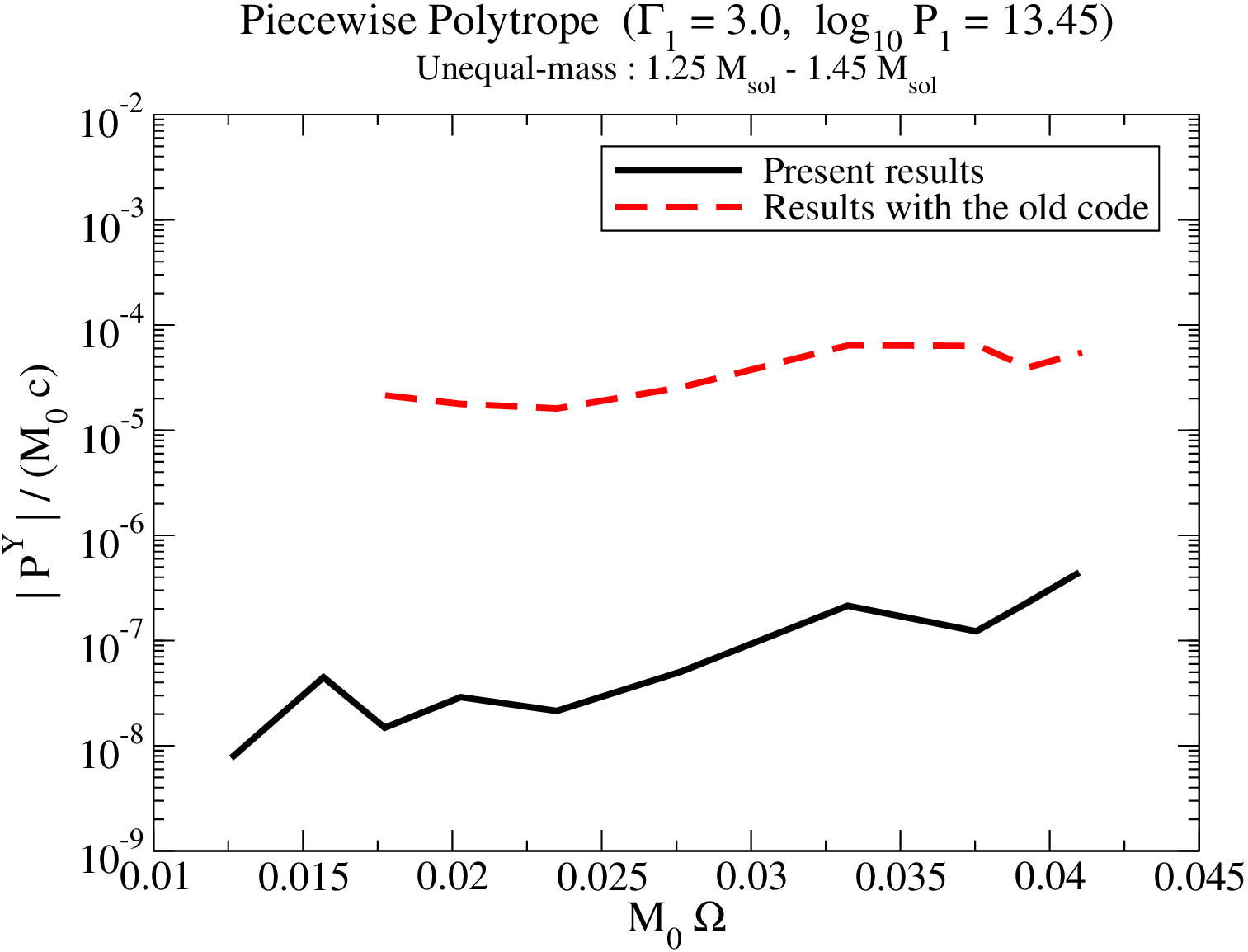}
\caption{Comparison of the linear momentum along a quasi-equilibrium
  sequence of an unequal-mass binary composed of
  $M_{\rm ADM}^{\rm NS1}=1.25 M_{\odot}$ and
  $M_{\rm ADM}^{\rm NS2}=1.45 M_{\odot}$ stars. The EOS we select is
  the piecewise polytrope with $\Gamma_1=3.0$ and
  $\log_{10} P_1=13.45$ (PwPoly30-1345). The thick (black) solid curve
  denotes the results calculated by our new code while the thick (red)
  dashed one is those calculated by the old code.
\label{fig7}}
\end{figure}

\subsection{Comparison with the results obtained by the old code}

We compare the linear momentum of an unequal-mass binary neutron
star along a quasi-equilibrium sequence for a model of piecewise
polytrope (PwPoly30-1345) with masses of
$M_{\rm ADM}^{\rm NS1}=1.25 M_{\odot}$ and
$M_{\rm ADM}^{\rm NS2}=1.45 M_{\odot}$ in Figure \ref{fig7}. This
figure compares the linear momentum of the $Y$-component, where we
assume that the centers of mass of the neutron stars are located on
the $X$-axis (see Figure \ref{fig6} about the location of each star).
This shows that the present results give by more than two orders
smaller values than those calculated by the old code through the
sequence. This improvement results from the change in the
solution method of the center of mass for achieving a convergence.
Note that the linear momentum of the $X$- and $Z$-directions is zero
within the machine precision because of the imposed symmetries.

\begin{figure}[ht]
\epsscale{1.0}
\plotone{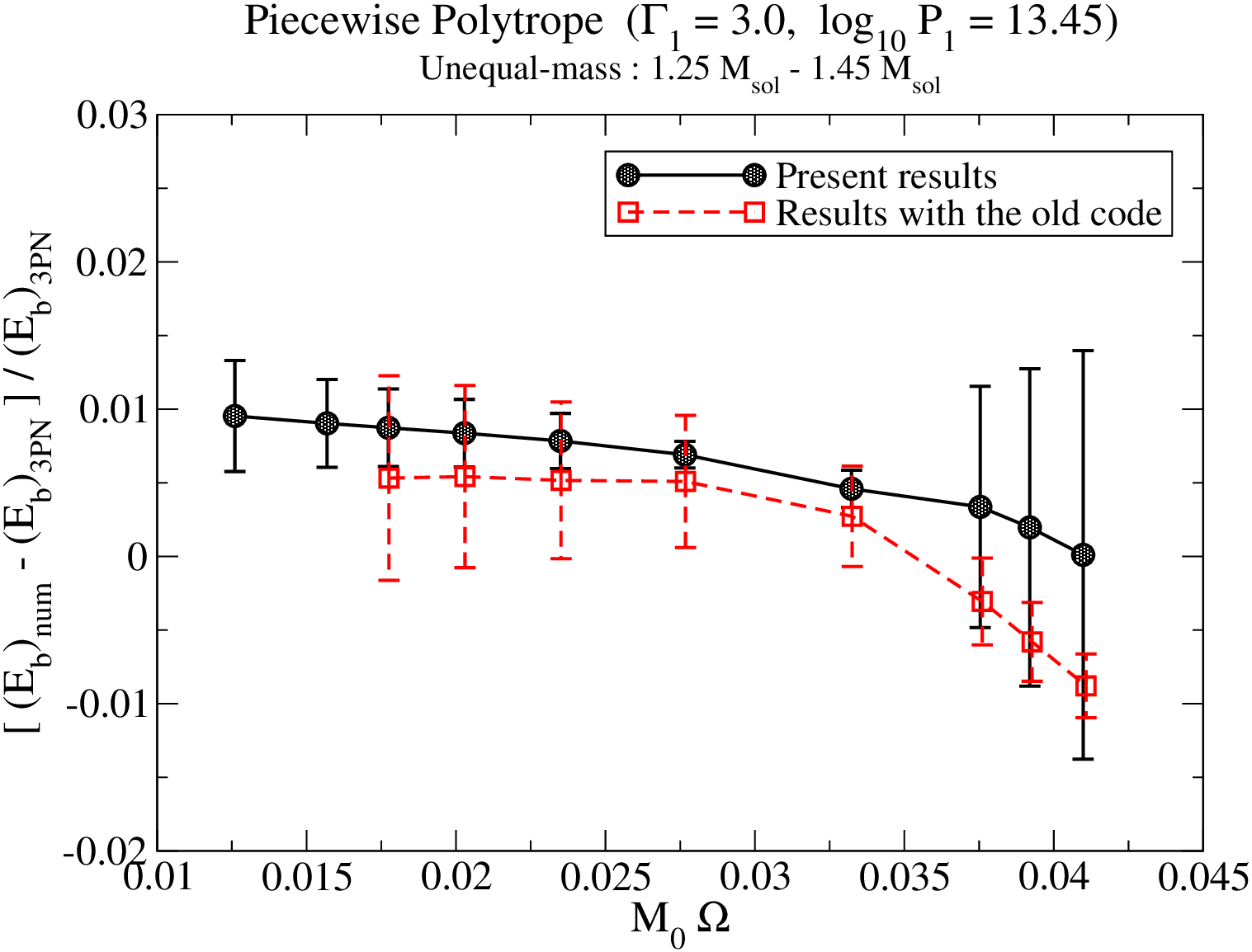}
\caption{Comparison of the relative difference of the numerically
  obtained binding energy from that by the 3PN approximation
  along the same quasi-equilibrium sequence in Figure \ref{fig7}. The
  (black) solid curve with filled circles denotes the relative
  difference of the binding energy calculated by our new code, and the
  (red) dashed curve with open squares is that computed by the old
  code. The error bar is drawn for an estimated error size derived
  from the virial error.
\label{fig8}}
\end{figure}

We also compare the relative difference of the binding energy from
that obtained by the 3PN approximation in Figure \ref{fig8}. The
definition of the relative difference is shown in
Equation (\ref{eq:deb}). The (black) solid curve with the filled circles
denotes the results calculated by the new code, and the (red) dashed
curve with the open squares is those computed by the old code. The
error bar is drawn for an estimated error size derived from the virial
error. It is found from Figure \ref{fig8} that the results by the new
code have a factor of 2--5 smaller error bar than those by the old
code except for very close separations ($M_0 \Omega > 0.037$).
Additionally, when we used the old code, we could not compute
sufficiently converged figures for larger separations ($M_0 \Omega <
0.017$), because the code fails to correctly determine the center of
mass during the computational iterations. For smaller mass ratio than
the model shown in Figure \ref{fig8}, $M_{\rm ADM}^{\rm NS 1}/M_{\rm
ADM}^{\rm NS 2} \simeq 0.862$, the old code also fails to achieve the
convergence, because the code crushes at earlier steps
of computational iteration. 

We conclude that our new code gives results as accurate as those
obtained by the old code. In addition, it can compute models with
smaller mass ratios that the old code cannot. 

\begin{figure}[t]
\epsscale{1.0}
\plotone{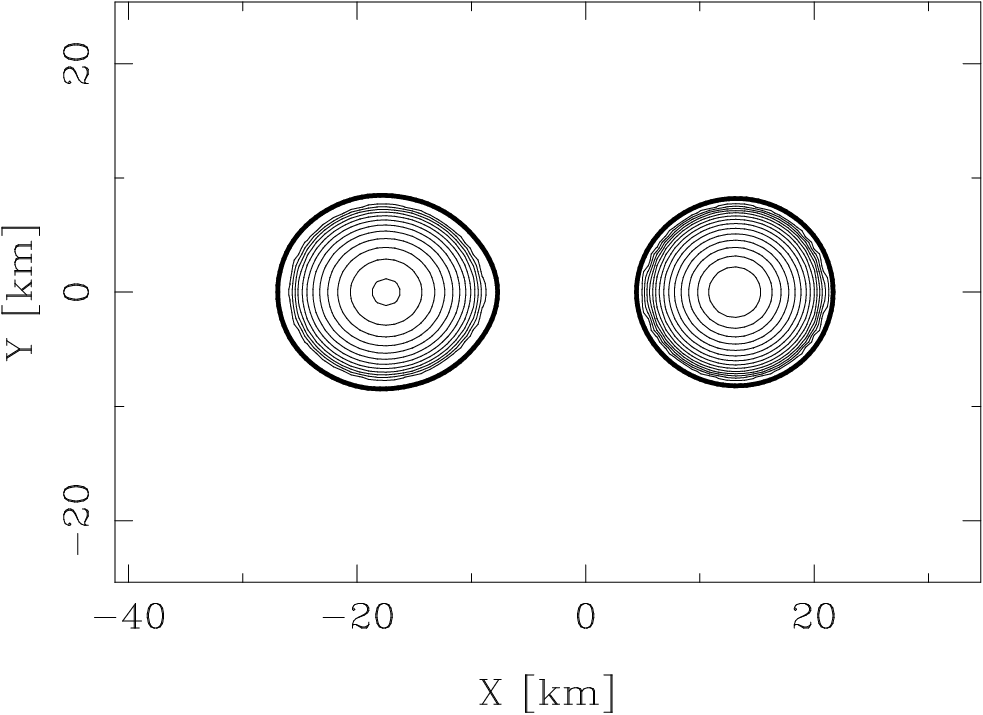}
\caption{Contours of baryon rest-mass density for the EOS of APR. 
  The star on the left-hand side has
  $M_{\rm ADM}^{\rm NS1}=1.15 M_{\odot}$ while that on the right-hand
  side does $M_{\rm ADM}^{\rm NS2}=1.55 M_{\odot}$. The thick solid
  circles denote the location of stellar surface. The axes are the
  coordinate length in km units.
\label{fig9}}
\end{figure}

\begin{figure}[ht]
\epsscale{1.0}
\plotone{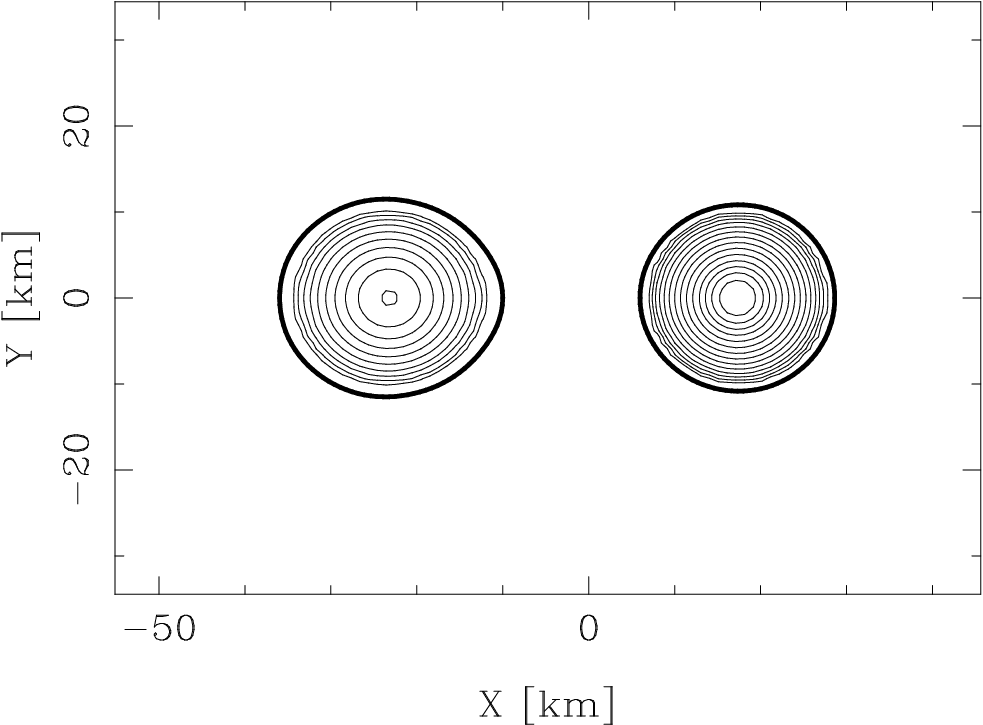}
\caption{Same as Figure \ref{fig9} but for the EOS of GNH3.
\label{fig10}}
\end{figure}

\section{Numerical Results} \label{sec:results}

Quasi-equilibrium sequences for 18 EOSs are computed, choosing three
total masses, $M_0=2.4 M_{\odot}$, $2.7 M_{\odot}$, and
$3.0 M_{\odot}$, and three mass ratios for each total mass. The
computation is performed with the collocation points of
$N_r \times N_{\theta} \times N_{\phi}=33 \times 25 \times 24$ for
larger separations and $33 \times 17 \times 16$ for closer ones. The
number of domains which cover the computational region around each
star is six for larger separations and five for closer ones. The
results are summarized in Appendix \ref{app:sequence}.

\begin{figure}[ht]
\epsscale{1.0}
\plotone{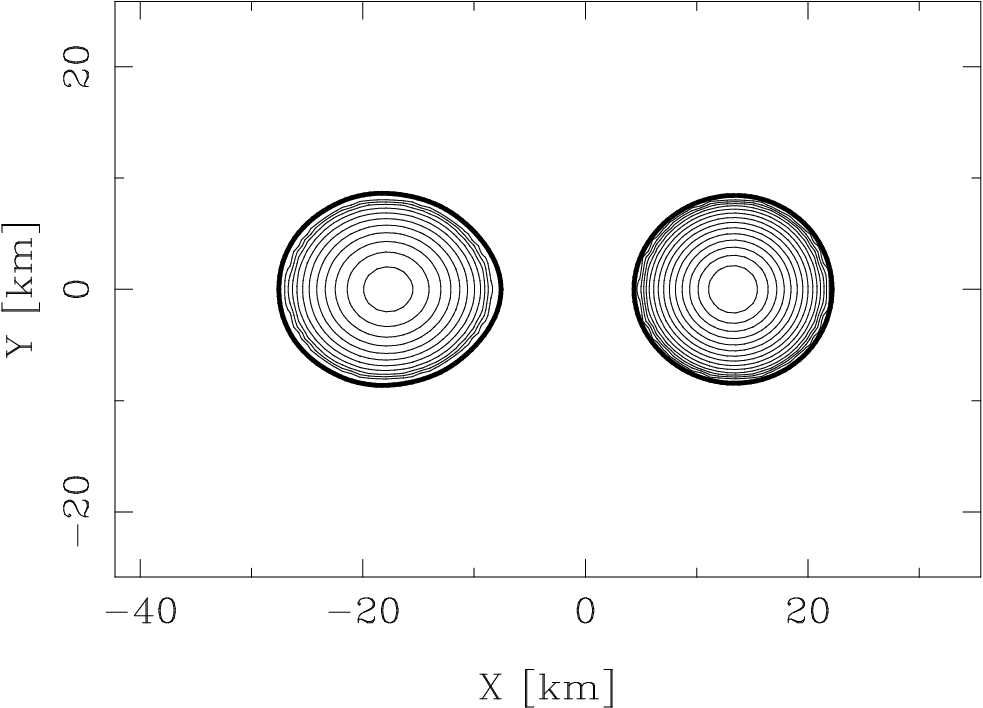}
\caption{Same as Figure \ref{fig9} but for the EOS of
  PwPoly30-1345.
\label{fig11}}
\end{figure}

\begin{figure}[ht]
\epsscale{1.0}
\plotone{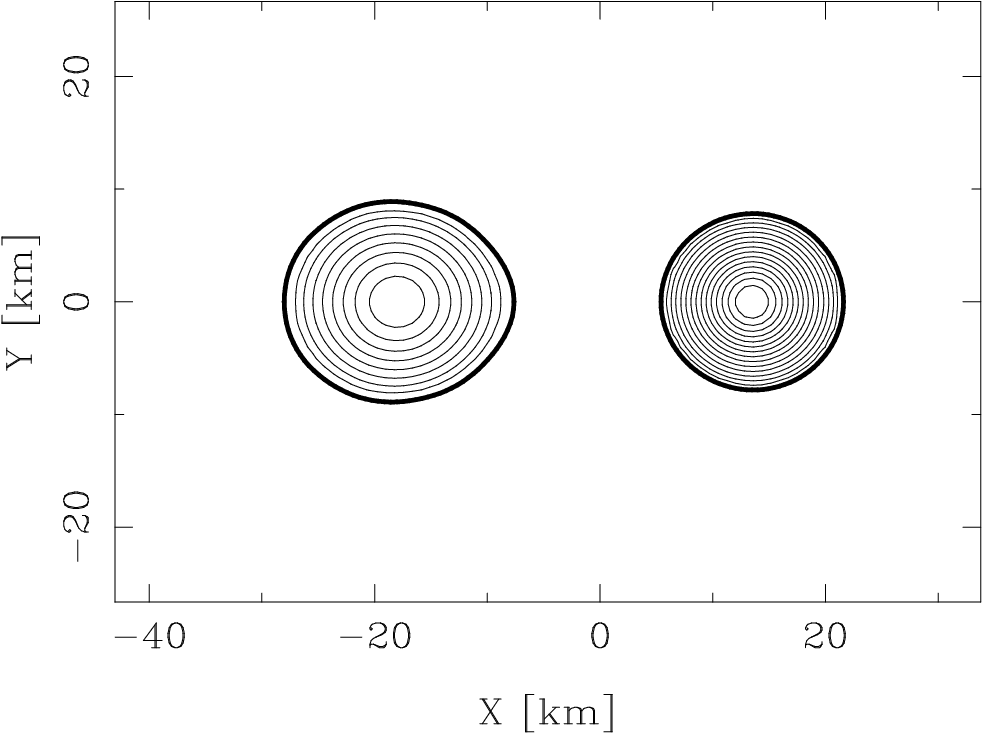}
\caption{Same as Figure \ref{fig9} but for the EOS of
  PwPoly24-1345.
\label{fig12}}
\end{figure}

\subsection{Contours of baryon rest-mass density}

Figures \ref{fig9}--\ref{fig12} show contours of baryon rest-mass
density for 4 EOSs in the equatorial plane. Masses of the stars on the
left- and right-hand sides are $M_{\rm ADM}^{\rm NS1}=1.15 M_{\odot}$
and $M_{\rm ADM}^{\rm NS2}=1.55 M_{\odot}$, respectively. All the
figures are drawn for the closest separation that we can choose. The
$X$- and $Y$-axes are measured by the coordinate length in km
units. Figure \ref{fig9} is for the tabulated EOS of APR and
Figure \ref{fig10} is for that of GNH3. The former EOS gives relatively
compact stars, while the latter one produces rather less compact
stars. Figures \ref{fig11} and \ref{fig12} are selected among the EOSs
of piecewise polytrope, PwPoly30-1345 and PwPoly24-1345. Both of the
piecewise polytropes produce stars of a similar size for a given mass,
$M_{\rm ADM}^{\rm NS} \simeq 1.3 M_{\odot}$,
but the former model, PwPoly30-1345, has a stiff EOS for the core
($\Gamma_1=3.0$) while the latter one, PwPoly24-1345, has a soft EOS
($\Gamma_1=2.4$). This difference in $\Gamma_1$ is reflected strongly
in the structure of more massive star for which the central density is
high and the effect of $\Gamma_1$ is appreciated. By contrast, a
striking difference is not seen for the less massive star.

\subsection{Binding energy and total angular momentum}

\begin{figure}[ht]
\epsscale{1.0}
\plotone{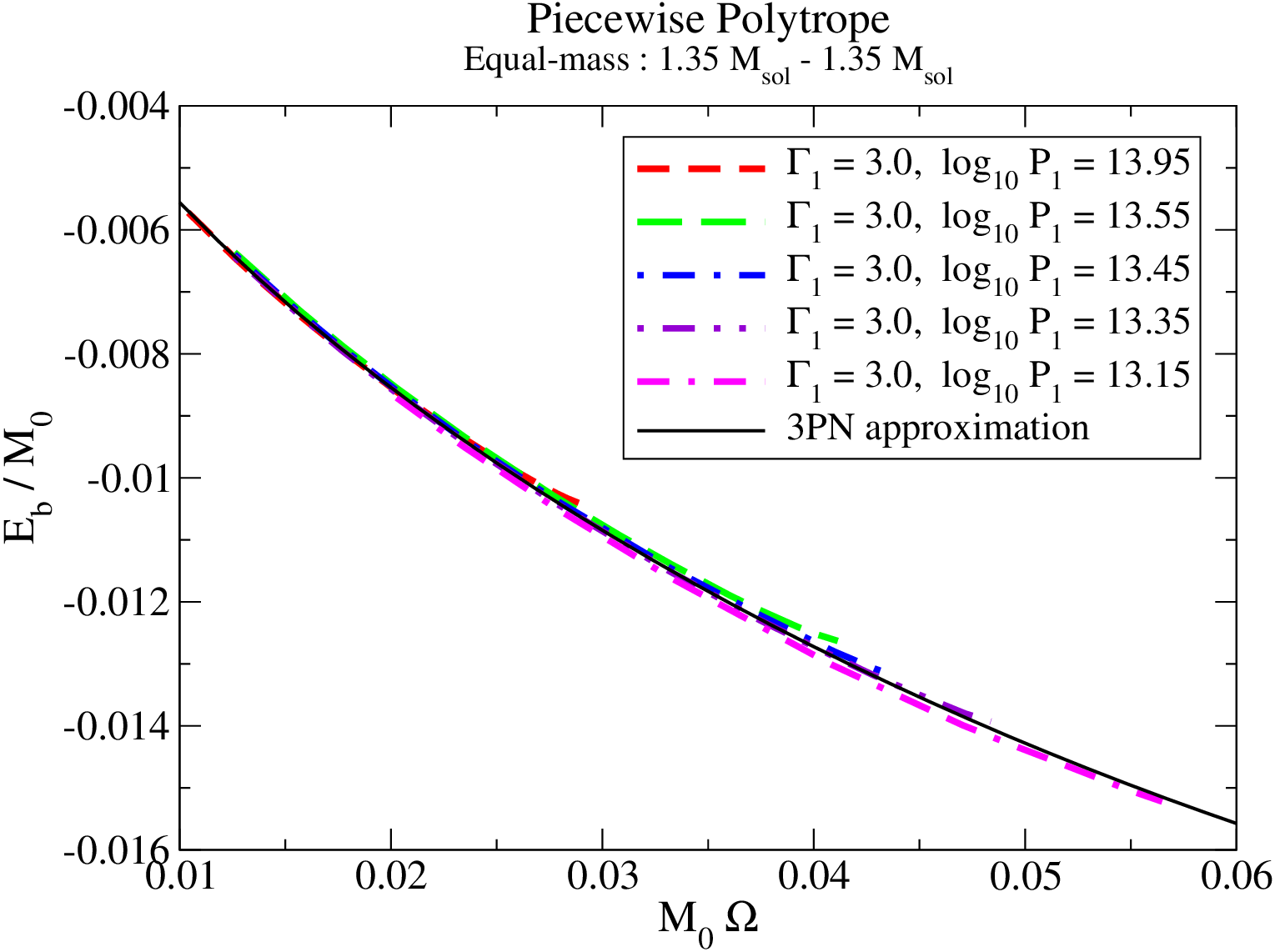}
\caption{Binding energy along a quasi-equilibrium sequence of an
  equal-mass binary neutron star as a function of the orbital angular
  velocity. Each ADM mass of the neutron stars is set to
  $1.35 M_{\odot}$ at infinite separation. The EOSs are the piecewise
  polytrope with $\Gamma_1=3.0$. The thick (red) short-dashed, thick
  (green) long-dashed, thick (blue) dot-dashed, thick (violet)
  dot-dot-dashed, and thick (magenta) dot-dash-dashed curves are,
  respectively, the cases of $\log_{10} P_1=13.95$ (PwPoly30-1395),
  13.55 (PwPoly30-1355), 13.45 (PwPoly30-1345), 13.35 (PwPoly30-1335),
  and 13.15 (PwPoly30-1315). The thin (black) solid curve denotes the
  results of the 3PN approximation.
\label{fig13}}
\end{figure}

\begin{figure}[ht]
\epsscale{1.0}
\plotone{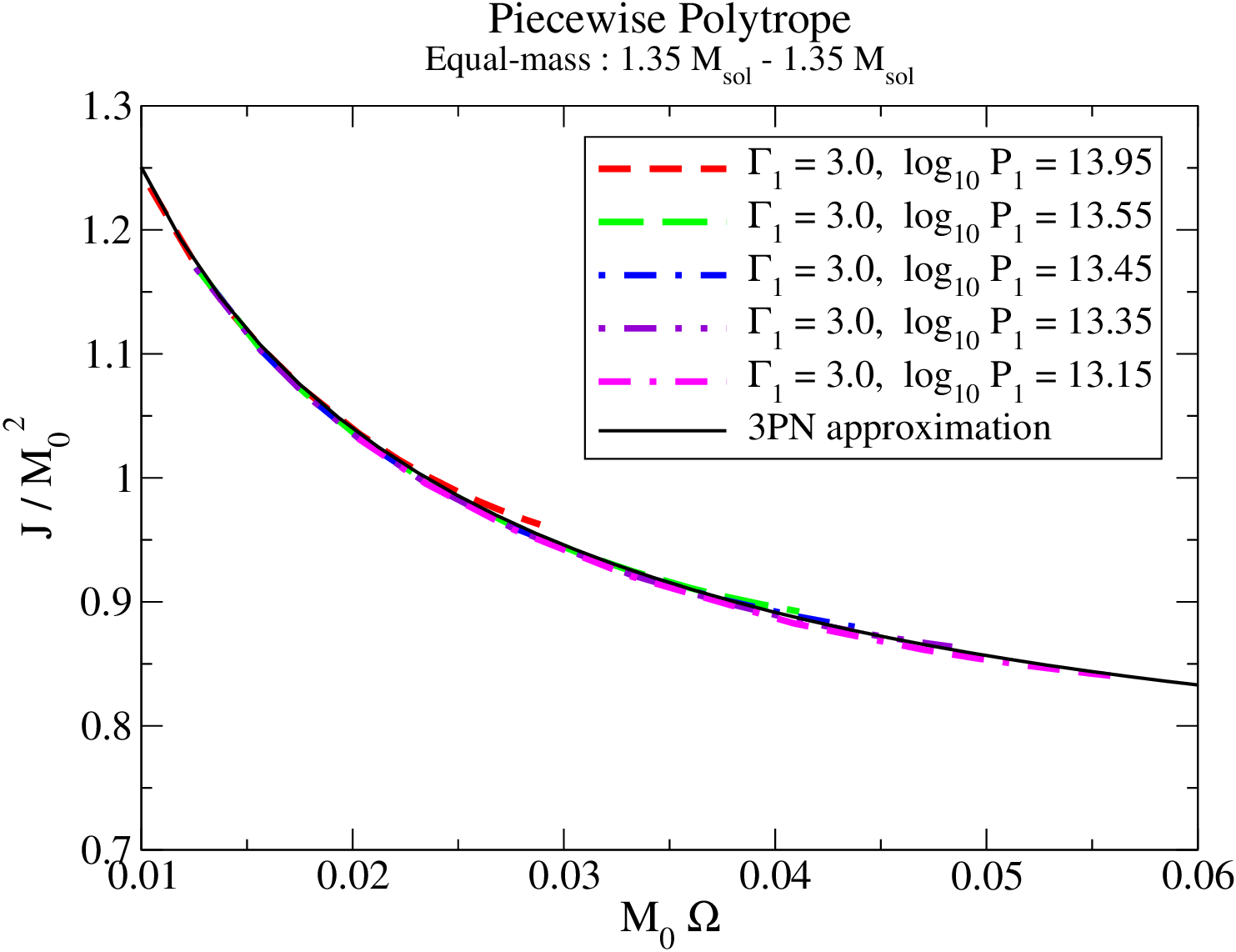}
\caption{Same as Figure \ref{fig13} but for the total angular momentum.
\label{fig14}}
\end{figure}

Figure \ref{fig13} shows the binding energy of quasi-equilibrium
sequences for five piecewise polytropic EOSs. Those sequences are
calculated for binary neutron stars composed of equal-mass stars of
$M_{\rm ADM}^{\rm NS1}=M_{\rm ADM}^{\rm NS2}=1.35 M_{\odot}$. All of
the piecewise polytropes we select here have $\Gamma_1=3.0$, but the
value of $\log_{10} P_1$ varies from 13.95 to 13.15. The thick (red)
short-dashed, thick (green) long-dashed, thick (blue) dot-dashed,
thick (violet) dot-dot-dashed, and thick (magenta) dot-dash-dashed
curves denote, respectively, the results for $\log_{10} P_1=13.95$
(PwPoly30-1395), 13.55 (PwPoly30-1355), 13.45 (PwPoly30-1345), 13.35
(PwPoly30-1335), and 13.15 (PwPoly30-1315). The thin (black) solid
curve denotes the results in the 3PN approximation. The total angular
momentum for the same EOSs is also plotted in Figure \ref{fig14}. The
sequences are terminated just before the stars reach the mass-shedding
limit, because the spectral method we use has a problem in handling a
cusp-like figure. We will discuss the endpoint in detail in
Section \ref{sec:endpoint}.

Figure \ref{fig13} shows that the orbital angular velocity at
the closest separation increases from $M_0 \Omega \simeq 0.029$ to
$\simeq 0.056$ as the value of $\log_{10} P_1$ decreases from 13.95
to 13.15. Because the model with $\log_{10} P_1=13.95$ (PwPoly30-1395)
has the largest radius for a spherical neutron star among the models
in Figure \ref{fig13}, the effect of tidal deformation is the
largest. We understand this behavior with the help of a Newtonian
analytic estimation as follows: by equating the gravity of a neutron
star (NS1) attracting on a test mass on the neutron star's surface
with the tidal force of the companion neutron star (NS2) attracting on
the test mass, we obtain the separation at which the mass-shedding
will occur for NS1. The separation is written in the form
\begin{equation}
  \frac{d_{\rm ms}}{r_{\rm NS1}} = A
  \Bigl( \frac{M_{\rm NS2}}{M_{\rm NS1}} \Bigr)^{1/3},
\end{equation}
where $A$, $r_{\rm NS1}$, $M_{\rm NS1}$, and $M_{\rm NS2}$ denote a
constant of order unity, the NS1's radius, the NS1's mass, and the
NS2's mass, respectively. Eliminating the separation at the 
mass-shedding from the Keplerian angular velocity
$\Omega = \sqrt{(M_{\rm NS1}+M_{\rm NS2})/d_{\rm ms}^3}$, we have
\begin{equation}
  M_{\rm tot} \Omega =\Bigl( \frac{M_{\rm tot}}{A r_{\rm NS1}}
  \Bigr)^{3/2} \Bigl( \frac{M_{\rm NS1}}{M_{\rm NS2}} \Bigr)^{1/2},
  \label{eq:omega_tid}
\end{equation}
where $M_{\rm tot} \equiv M_{\rm NS1}+M_{\rm NS2}$ is the total mass.
Taking the ratio of the orbital angular velocity at the mass-shedding 
for PwPoly30-1395 to that for PwPoly30-1315, we obtain
\begin{equation}
  \frac{(M_{\rm tot} \Omega)_{1395}}
       {(M_{\rm tot} \Omega)_{1315}}
       =\Bigl( \frac{r_{\rm NS1}^{1315}}
       {r_{\rm NS1}^{1395}} \Bigr)^{3/2}.
\end{equation}
Noting that the masses are $M_{\rm NS1}=M_{\rm NS2}=1.35 M_{\odot}$
and $r_{\rm NS1}$ is the circumferential radius, we have the ratio as
0.51 from Table \ref{table2}. This value agrees approximately with the
obtained ratio of the orbital angular velocity at the closest
separation,
\begin{equation}
  \frac{(M_0 \Omega)_{1395}}{(M_0 \Omega)_{1315}} \simeq
    \frac{0.029}{0.056} \simeq 0.52.
\end{equation}

As a result of the smaller orbital angular velocity at the closest
separation, the model with $\log_{10} P_1=13.95$ (PwPoly30-1395) has
the nondimensional angular momentum as large as $J/M_0^2 \sim 0.96$,
as shown in Figure \ref{fig14}. By contrast, the most compact model with
$\log_{10} P_1=13.15$ (PwPoly30-1315) has $J/M_0^2 \sim 0.84$, much
smaller than 0.96, because the binary system can come closer than that
composed of less compact stars. These differences suggest that the
merger process will depend strongly on the EOS: for the stiffer EOS,
the nondimensional angular momentum at the onset of merger may be
too large to form a black hole soon, whereas for the softer EOS, it
may be small enough that the merged object collapses to a black hole
in a dynamical timescale $\sim 1$ ms.

\begin{figure}[ht]
\epsscale{1.0}
\plotone{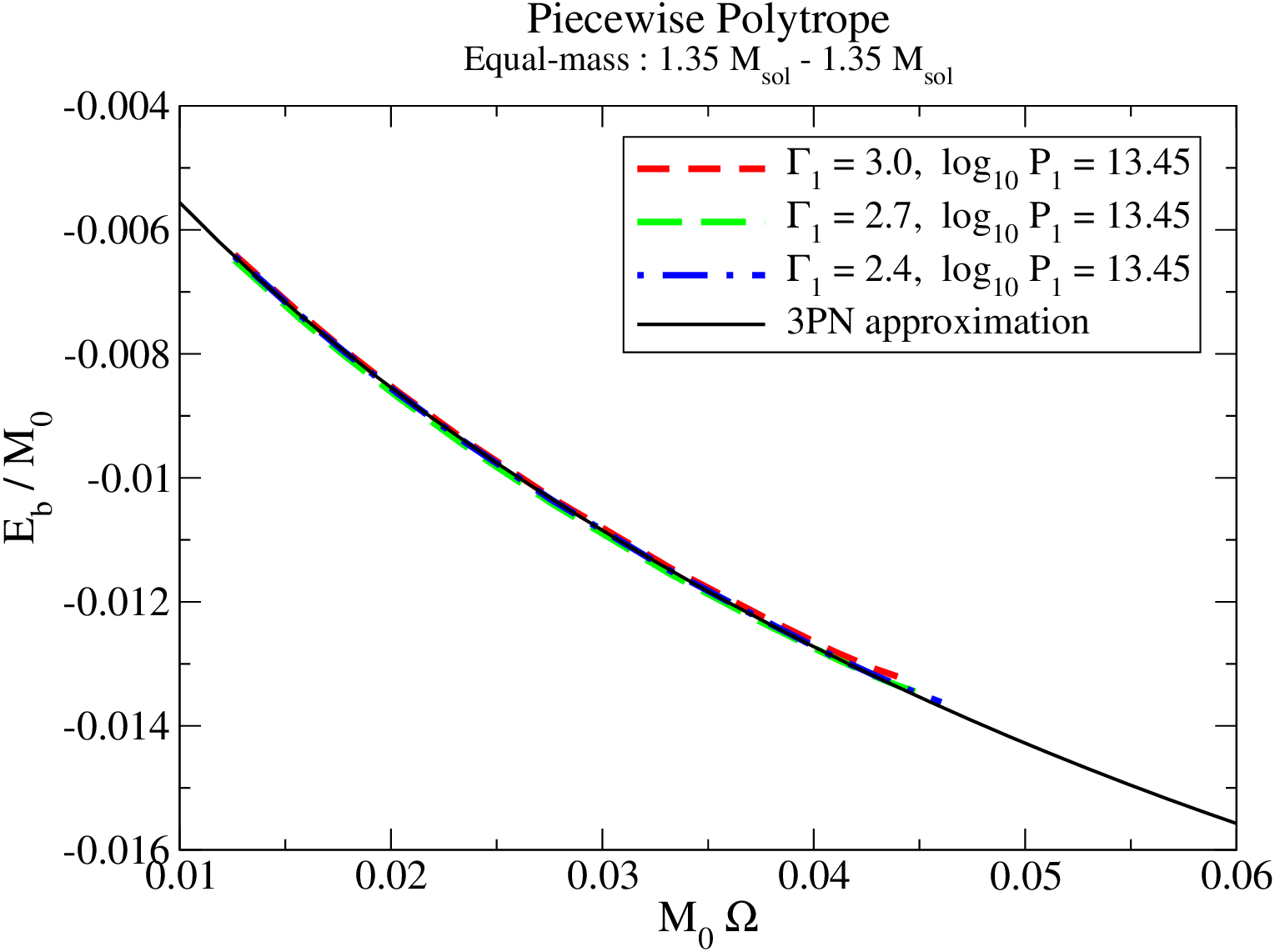}
\caption{Same as Figure \ref{fig13} but for the piecewise polytrope with
  $\log_{10} P_1=13.45$. The thick (red) short-dashed, thick (green)
  long-dashed, and thick (blue) dot-dashed curves denote the cases
  with the adiabatic index of $\Gamma_1=3.0$ (PwPoly30-1345), 2.7
  (PwPoly27-1345), and 2.4 (PwPoly24-1345), respectively. The thin
  (black) solid curve denotes the results of the 3PN approximation.
\label{fig15}}
\end{figure}

Figure \ref{fig15} is drawn for the binding energy of equal-mass
binary neutron stars with $M_{\rm ADM}^{\rm NS1}=M_{\rm ADM}^{\rm
NS2}=1.35 M_{\odot}$. The chosen EOSs are the piecewise polytrope with
a fixed value of $\log_{10} P_1=13.45$, but the value of $\Gamma_1$ is
varied from 3.0 to 2.4. The thick (red) short-dashed, thick (green)
long-dashed, and thick (blue) dot-dashed curves denote the results for
$\Gamma_1=3.0$ (PwPoly30-1345), 2.7 (PwPoly27-1345), and 2.4
(PwPoly24-1345), respectively. For these three models, the radius of a
spherical star for $M_{\rm ADM}^{\rm NS}=1.35 M_{\odot}$ has almost
the same value but very slightly more compact for $\Gamma_1=2.4$ than
for $\Gamma_1=3.0$ (see Figure \ref{fig1} and Table \ref{table2}). It is
found that the orbital angular velocity at the closest separation
increases from $M_0 \Omega \simeq 0.044$ to 0.046 as the adiabatic
index of the neutron star core is decreased from $\Gamma_1=3.0$ to
2.4. This variation of the orbital angular velocity is smaller than
that by the change of $\log_{10} P_1$ as seen in
Figure \ref{fig13}. This implies that for neutron stars of mass
$1.35M_{\odot}$, the variation in $\log_{10} P_1$, which affects the
stellar radius, results in a larger effect on the determination of the
orbital angular velocity at the closest separation than the effect of
$\Gamma_1$. The reason for this is that the maximum density for the
neutron star mass of $1.35M_{\odot}$ is not so large that their
structure depends weakly on $\Gamma_1$, which determines the stiffness
of the core EOS.

\begin{figure}[ht]
\epsscale{1.0}
\plotone{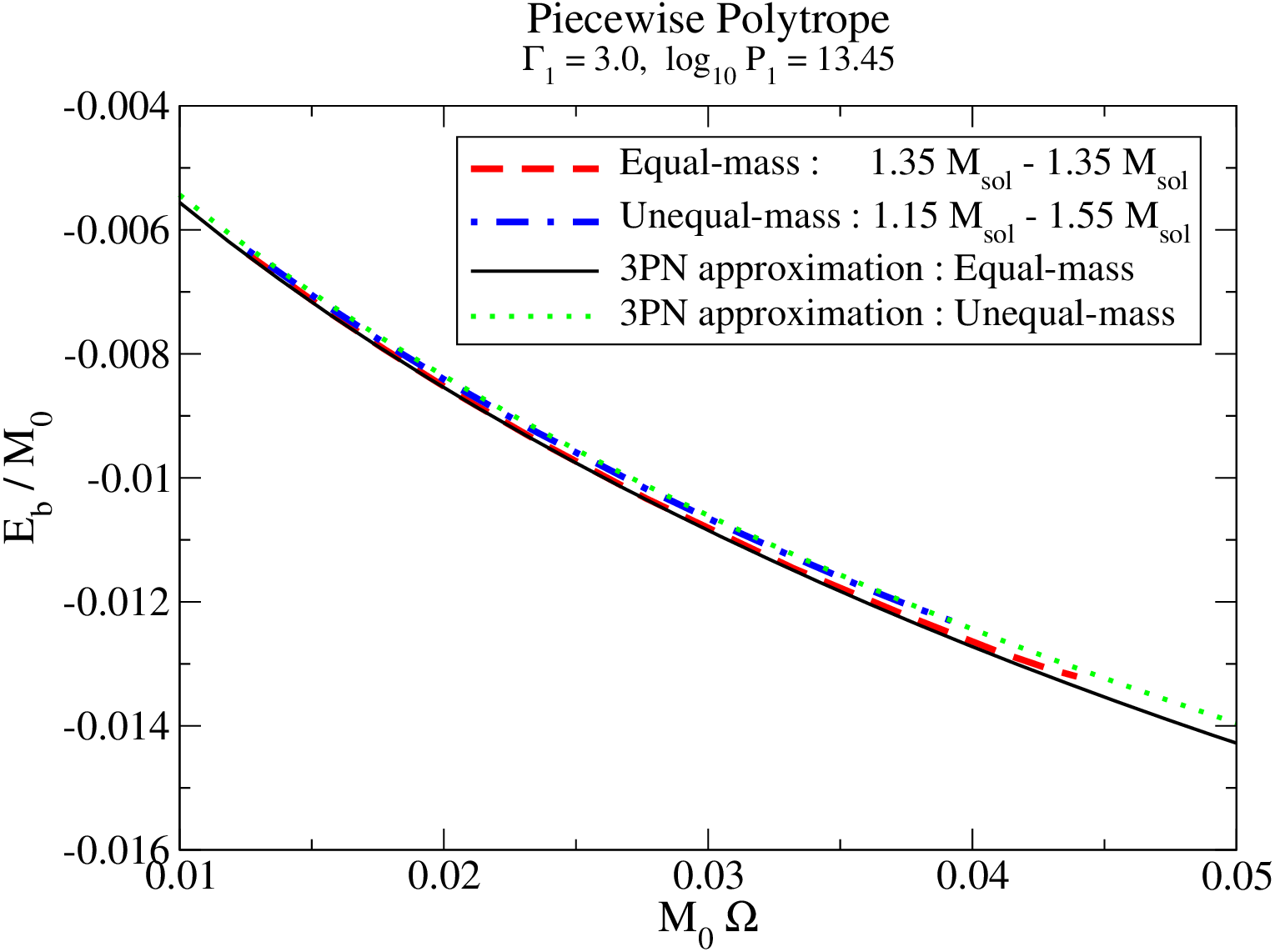}
\caption{Same as Figure \ref{fig13} but for comparing the results of
  equal-mass binary neutron stars with those of unequal-mass ones.
  The EOS is the piecewise polytrope with $\Gamma_1=3.0$ and
  $\log_{10} P_1=13.45$ (PwPoly30-1345). The thick (red) dashed curve
  denotes the case of an equal-mass binary with
  $M_{\rm ADM}^{\rm NS1}=M_{\rm ADM}^{\rm NS2}=1.35 M_{\odot}$, while
  the thick (blue) dot-dashed one is the case of an unequal-mass
  binary composed of $M_{\rm ADM}^{\rm NS1}=1.15 M_{\odot}$ and
  $M_{\rm ADM}^{\rm NS2}=1.55 M_{\odot}$ stars. The thin (black) solid
  and thin (green) dotted curves are the results of the 3PN
  approximation for equal-mass and unequal-mass binaries,
  respectively.
\label{fig16}}
\end{figure}

\begin{figure}[ht]
\epsscale{1.0}
\plotone{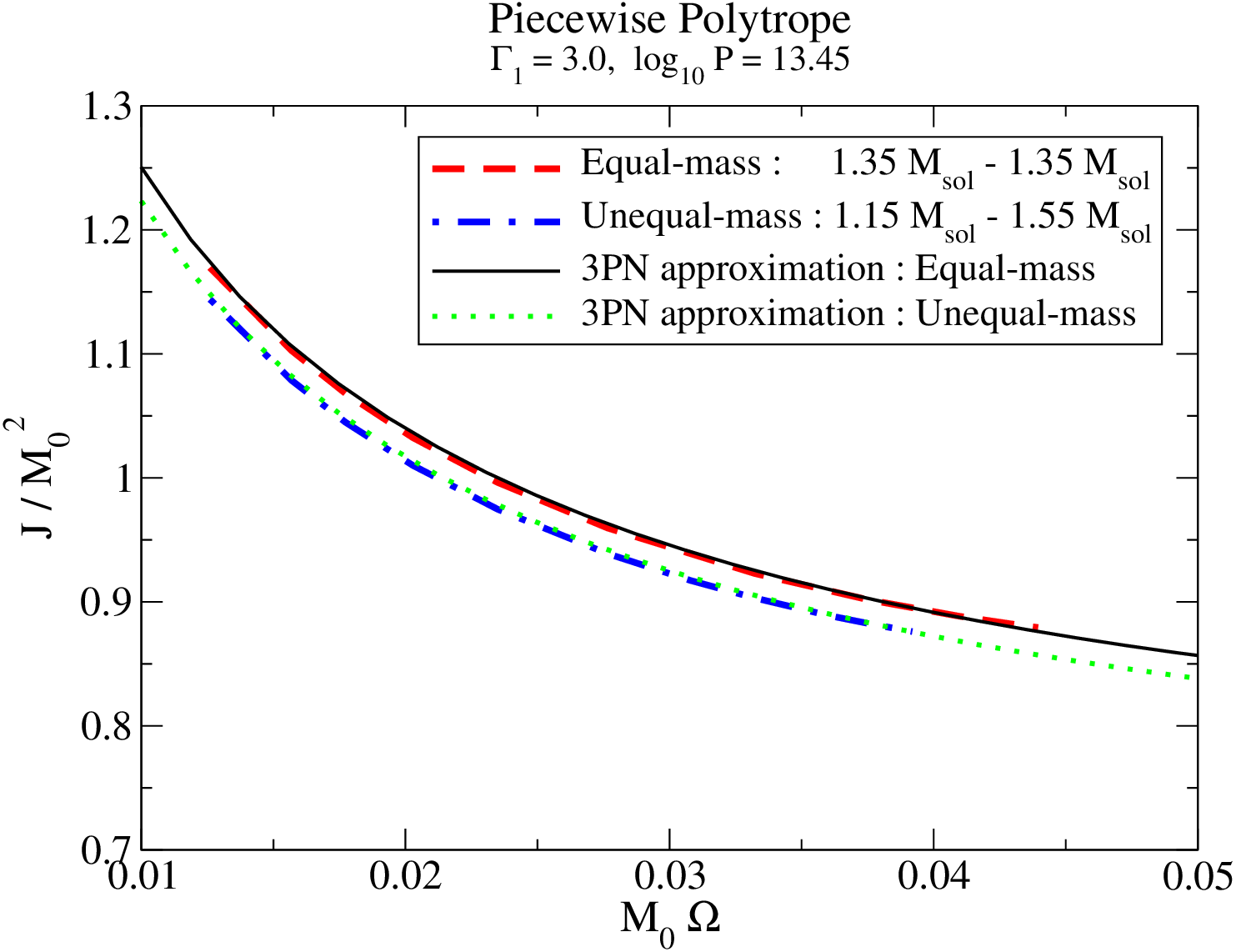}
\caption{Same as Figure \ref{fig16} but for the total angular momentum.
\label{fig17}}
\end{figure}

Figure \ref{fig16} compares the binding energy of an equal-mass binary
neutron star with that of an unequal-mass one. The EOS we choose here
is the piecewise polytrope with $\Gamma_1=3.0$ and $\log_{10}
P_1=13.45$ (PwPoly30-1345). The thick (red) dashed curve denotes the
result for an equal-mass binary with $M_{\rm ADM}^{\rm NS1}=M_{\rm
ADM}^{\rm NS2}=1.35 M_{\odot}$, while the thick (blue) dot-dashed one
is for an unequal-mass binary composed of $M_{\rm ADM}^{\rm NS1}=1.15
M_{\odot}$ and $M_{\rm ADM}^{\rm NS2}=1.55 M_{\odot}$ stars. The thin
(black) solid and thin (green) dotted curves are the results of the
3PN approximation for the equal-mass and unequal-mass binaries,
respectively. When the mass ratio decreases to the value as small as
$1.15 M_{\odot}/1.55 M_{\odot} \simeq 0.74$, the fractional binding
energy $E_{\rm b}/M_0$ and orbital angular velocity $M_0 \Omega$ at
the closest separation decrease by about 10\%. The reason why the
orbital angular velocity at the closest separation decreases for the
unequal-mass case is that the less massive star is tidally deformed by
the companion more massive star and starts shedding mass at a larger
separation than that for the equal-mass case. The binding energy
decreases for the unequal-mass case because it is proportional to the
reduced mass $\mu \equiv M_{\rm ADM}^{\rm NS1} M_{\rm ADM}^{\rm
NS2}/M_0$ according to the results of the 3PN approximation (the ratio
of the unequal-mass case to the equal-mass one is $\mu_{\rm
uneq}/\mu_{\rm eq} \simeq 0.978$) and the orbital angular velocity at
the termination point of the sequence should be smaller.

In Figure \ref{fig17}, the total angular momentum for the same models as
in Figure \ref{fig16} is shown along both equal-mass and unequal-mass
sequences. The sequence of the total angular momentum for the
unequal-mass case is located below that of the equal-mass case at the
same orbital angular velocity. This is also because the total angular
momentum is proportional to the reduced mass, according to the results
of the 3PN approximation. However, because the orbital angular
velocity at the termination point of the sequence is smaller for the
unequal-mass case, its total angular momentum is approximately the
same value as that for the equal-mass case coincidentally. 

\begin{figure}[ht]
\epsscale{1.0}
\plotone{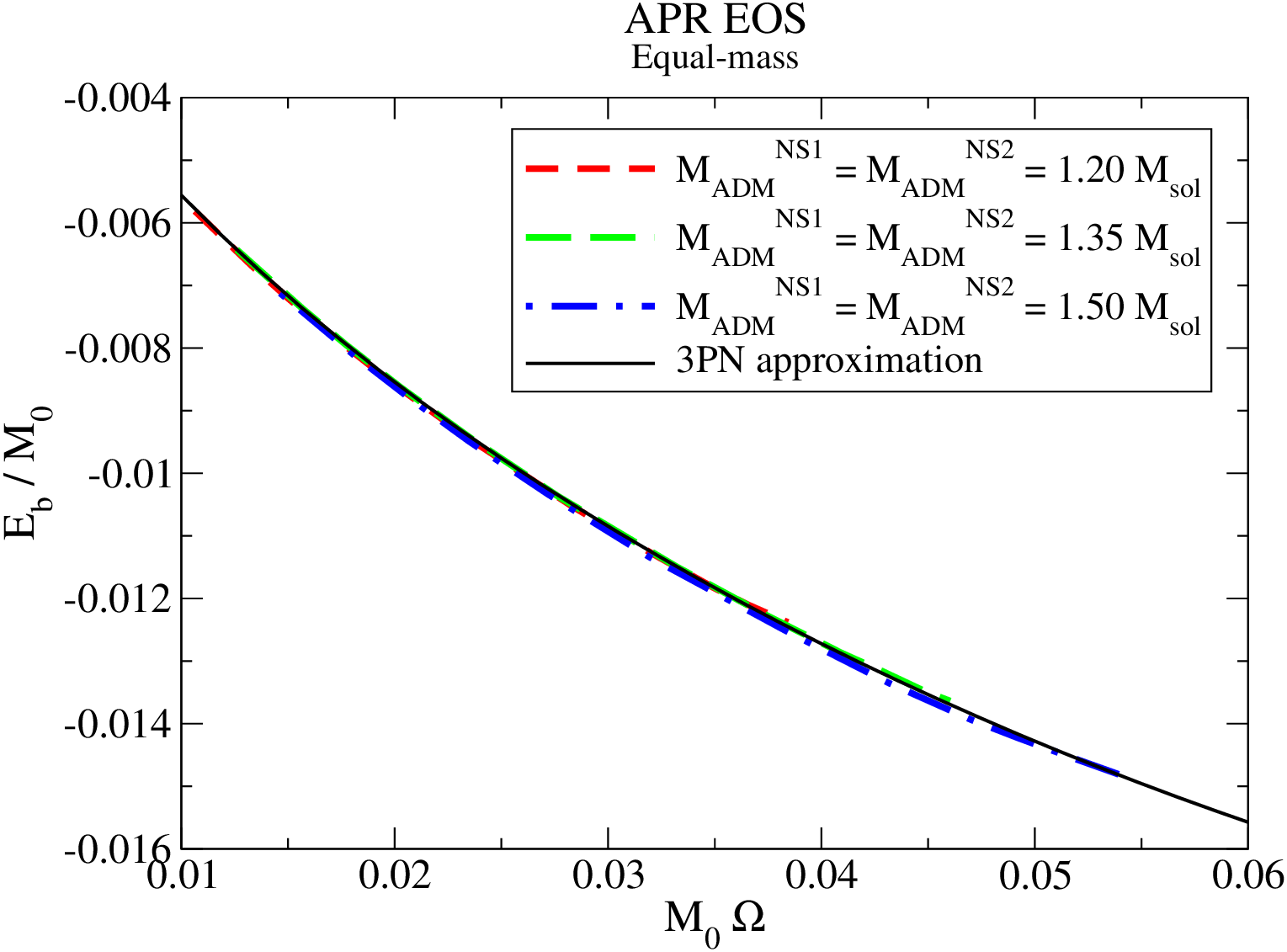}
\caption{Same as Figure \ref{fig13} but for the APR EOS. The thick (red)
  short-dashed, thick (green) long-dashed, and thick (blue) dot-dashed
  curves denote the results of equal-mass binary neutron stars with
  the total mass of $M_0=2.4 M_{\odot}$, $2.7 M_{\odot}$, and
  $3.0 M_{\odot}$, respectively. The thin (black) solid curve denotes the
  results of the 3PN approximation.
\label{fig18}}
\end{figure}

Finally, we show the effect of the total mass of binary neutron stars
on the binding energy in Figure \ref{fig18}. The EOS we choose for this
figure is one of the tabulated realistic EOSs, the APR EOS.  Figure
\ref{fig18} shows the results for three total masses, $2.4 M_{\odot}$,
$2.7 M_{\odot}$, and $3.0 M_{\odot}$. As the neutron star mass
increases, the star becomes more compact and less subject to tidal
disruption. For a binary system composed of more massive neutron
stars, the two stars are necessary to come closer each other to reach
their mass-shedding limit. This makes the orbital angular velocity at
the mass-shedding limit increase for binary systems with massive
neutron stars. Because of this effect, the orbital angular velocity at
the closest separation is about 0.038 for $M_0=2.4 M_{\odot}$ whereas
it is about 0.054 for $M_0=3.0 M_{\odot}$. We can understand this
behavior by using Equation (\ref{eq:omega_tid}). By substituting each total
mass and circumferential radius, we obtain the ratio of the orbital
angular velocity as
\begin{equation}
  \frac{(M_{\rm tot} \Omega)_{2.4 M_{\odot}}}
  {(M_{\rm tot} \Omega)_{3.0 M_{\odot}}}
  \simeq \Bigl( \frac{2.4 M_{\odot}/11.37~{\rm km}}
  {3.0 M_{\odot}/11.31~{\rm km}} \Bigr)^{3/2} \simeq 0.71.
  \label{eq:omega_ratio}
\end{equation}
The actual ratio of the orbital angular velocity at the closest
separation we obtained is 
\begin{equation}
  \frac{(M_0 \Omega)_{2.4 M_{\odot}}}
  {(M_0 \Omega)_{3.0 M_{\odot}}} \simeq
  \frac{0.038}{0.054} \simeq 0.70.
\end{equation}
This value agrees again with the estimated ratio given in 
Equation (\ref{eq:omega_ratio}).

\subsection{Endpoint of sequences} \label{sec:endpoint}

\begin{figure}[ht]
\epsscale{1.0}
\plotone{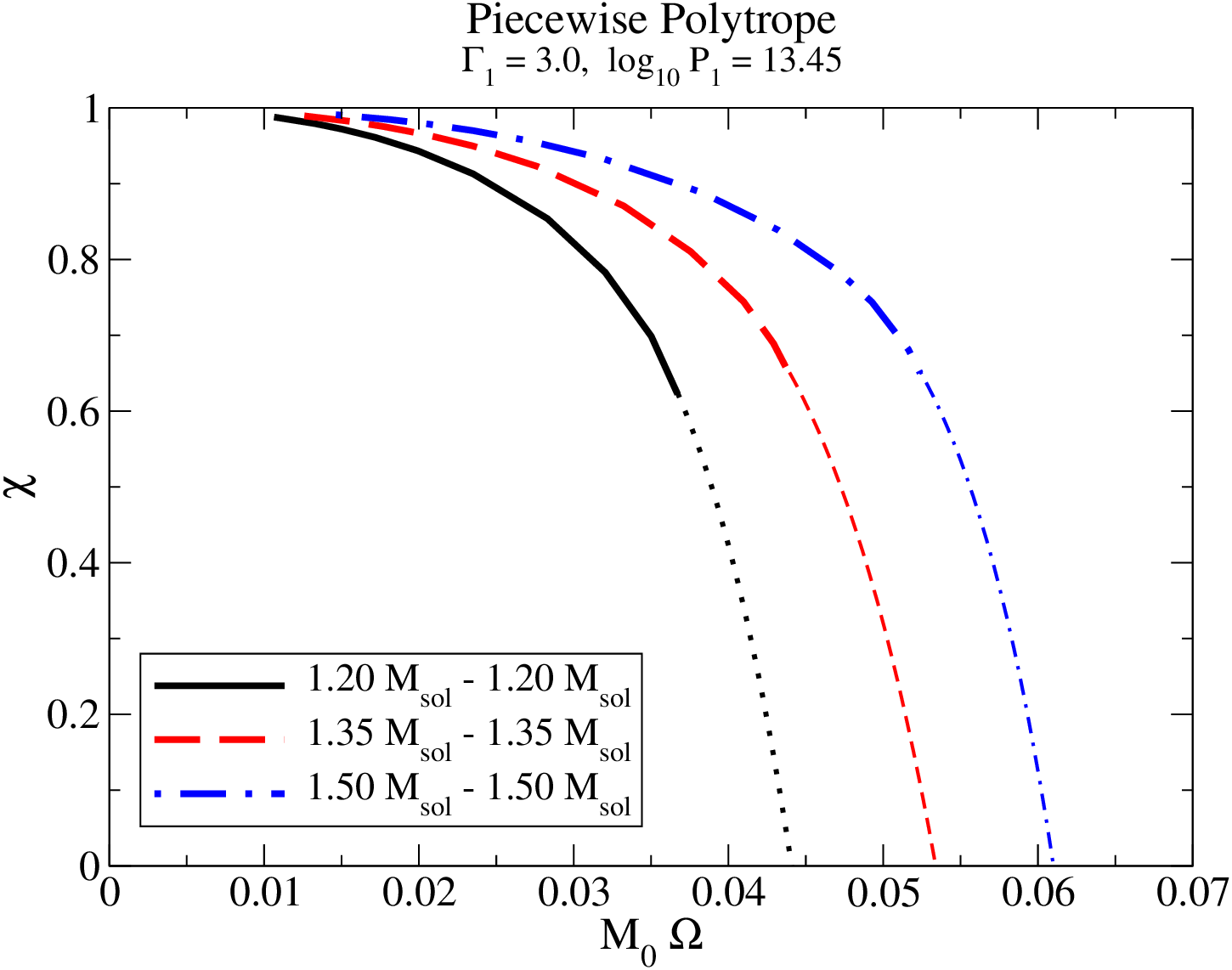}
\caption{Mass-shedding indicator $\chi$ as a function of the orbital
  angular velocity. The EOS we select is the piecewise polytrope with
  $\Gamma_1=3.0$ and $\log_{10} P_1=13.45$ (PwPoly30-1345), and
  sequences of the equal-mass case are shown. The thick (black) solid,
  thick (red) dashed, and thick (blue) dot-dashed curves denote the
  computed sequences for equal-mass binary neutron stars with masses
  of $M_{\rm ADM}^{\rm NS1}=M_{\rm ADM}^{\rm NS2}=1.20 M_{\odot}$,
  $1.35 M_{\odot}$, and $1.50 M_{\odot}$, respectively. The thin
  (black) dotted, thin (red) dashed, and thin (blue) dot-dashed curves
  denote the extrapolated curves.
\label{fig19}}
\end{figure}

We construct quasi-equilibrium sequences for three cases of total
mass, three mass ratios for each total mass, and 18 EOSs. For all
the sequences, we do not find the turning point of the binding energy
and total angular momentum, which represents the innermost stable
circular orbit of the binary system. Instead, the sequences terminate
at the mass-shedding limit of a less massive star. (For equal-mass
binaries, two neutron stars reach the mass-shedding limit at the same
time.) As explained before, we stop constructing the sequences at
$\chi \sim 0.6$ because our spectral method code cannot handle a
cusp-like figure which appears at the mass-shedding limit. This
implies that the closest separation with the largest value of
$M_0 \Omega$ we can calculate is not the actual endpoint of the
sequence.

\begin{figure}[ht]
\epsscale{1.0}
\plotone{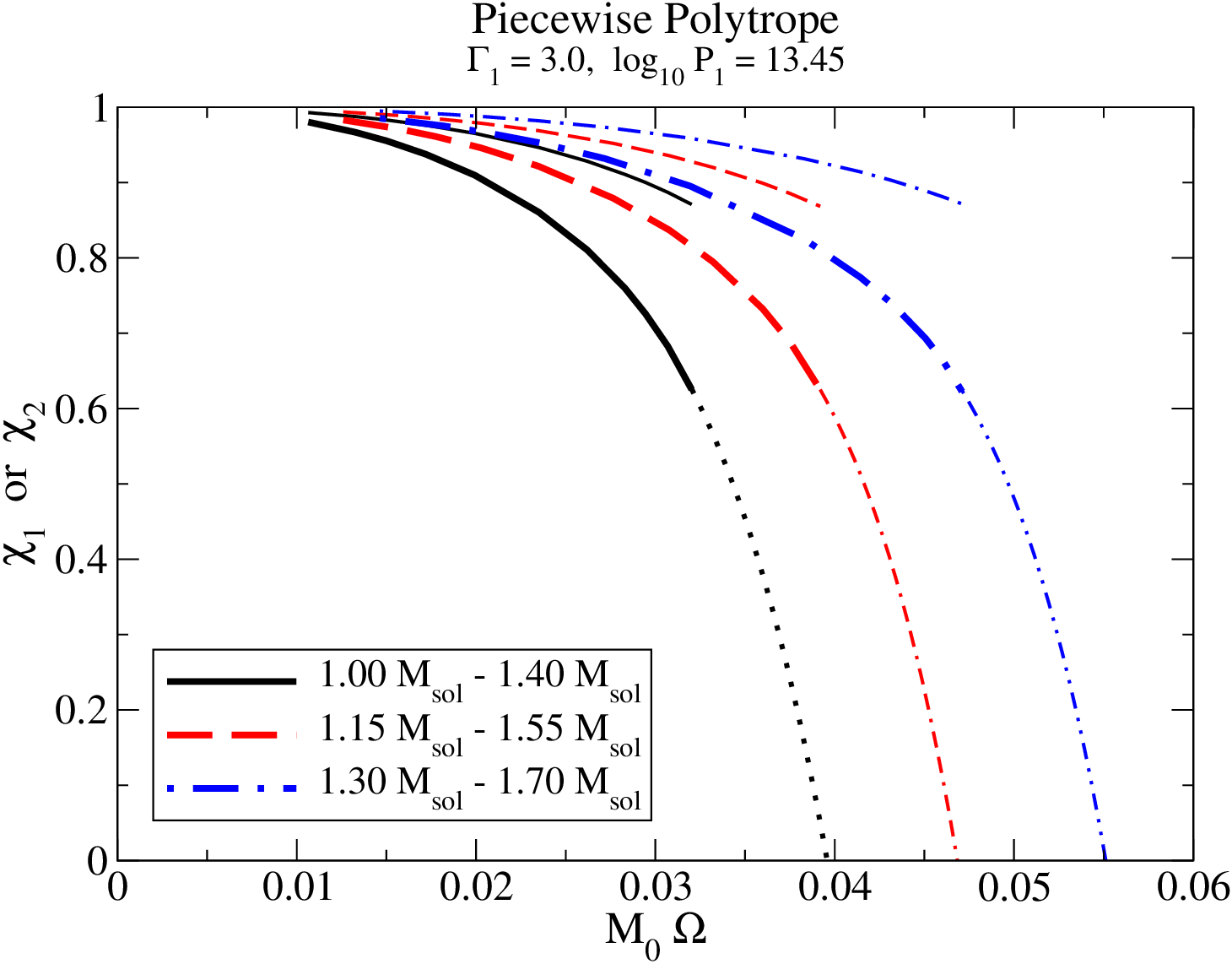}
\caption{Same as Figure \ref{fig19} but for unequal-mass models. The
  (black) solid, (red) dashed, and (blue) dot-dashed curves denote
  the computed sequences with masses of $M_{\rm ADM}^{\rm NS1}$ vs.
  $M_{\rm ADM}^{\rm NS2}$ $=1.00 M_{\odot}$ vs. $1.40 M_{\odot}$,
  $1.15 M_{\odot}$ vs. $1.55 M_{\odot}$, and $1.30 M_{\odot}$ vs.
  $1.70 M_{\odot}$, respectively. For each curve above, the thick and
  thin ones denote the sequences for less massive stars and those for
  more massive ones, respectively. The thin (black) dotted, thin (red)
  dot-dash-dashed, and thin (blue) dot-dot-dashed curves are the
  extrapolated ones.
\label{fig20}}
\end{figure}

To determine the orbital angular velocity at the mass-shedding limit,
we extrapolate a curve of the mass-shedding indicator $\chi$ as a
function of $M_0 \Omega$. The procedure is as follows: we first make a
fitting polynomial equation of $\chi$ as a function of $M_0 \Omega$
for sequences. Then, we extrapolate the fitting polynomial equation to
$\chi=0$ and determine the value of $M_0 \Omega$ at $\chi=0$.  The
extrapolated value of $M_0 \Omega$ at $\chi=0$ is defined as $M_0
\Omega_{\rm ms}$. Figure \ref{fig19} shows such extrapolated curves
for the piecewise polytropic EOS with $\Gamma_1=3.0$ and $\log_{10}
P_1=13.45$ (PwPoly30-1345). The thick (black) solid, thick (red)
dashed, and thick (blue) dot-dashed curves denote the computed
sequences for equal-mass binary neutron stars with masses of $M_{\rm
ADM}^{\rm NS1}=M_{\rm ADM}^{\rm NS2}=1.20 M_{\odot}$, $1.35
M_{\odot}$, and $1.50 M_{\odot}$, respectively. The thin (black)
dotted, thin (red) dashed, and thin (blue) dot-dashed curves denote
the extrapolated curves.

For unequal-mass binaries, we apply the same method of extrapolation
to less massive stars which will be tidally disrupted by their companion
more massive stars. Figure \ref{fig20} shows the sequences and
extrapolated curves for unequal-mass binary neutron stars for the
piecewise polytropic EOS with $\Gamma_1=3.0$ and $\log_{10} P_1=13.45$
(PwPoly30-1345). The (black) solid, (red) dashed, and (blue)
dot-dashed curves denote the computed sequences with masses of
$M_{\rm ADM}^{\rm NS1}$ versus $M_{\rm ADM}^{\rm NS2}$
$=1.00 M_{\odot}$ versus $1.40 M_{\odot}$, $1.15 M_{\odot}$ versus
$1.55 M_{\odot}$, and $1.30 M_{\odot}$ versus $1.70 M_{\odot}$,
respectively. For each curve, the thick and thin ones denote the
sequences for less massive stars and those for more massive ones,
respectively. The thin (black) dotted, thin (red) dot-dash-dashed, and
thin (blue) dot-dot-dashed curves are the extrapolated ones.

\begin{figure}[t]
\epsscale{1.0}
\plotone{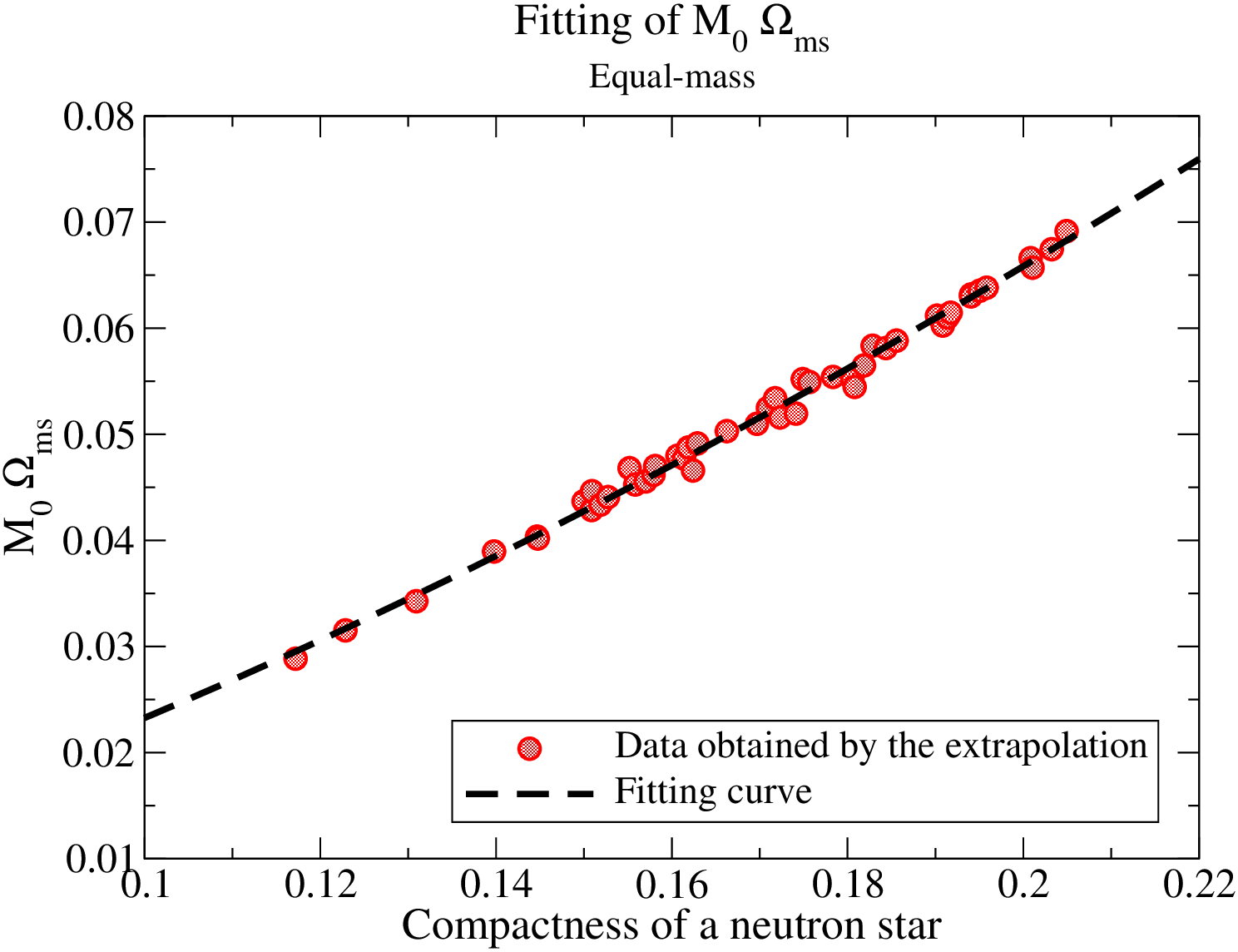}
\caption{Fitting curve for the equal-mass models. The filled (red)
  circles denote the orbital angular velocity at the mass-shedding
  limit obtained by our extrapolation method as a function of the
  compactness of a neutron star. The (black) dashed curve is the
  fitting curve for the data.
\label{fig21}}
\end{figure}

The estimated orbital angular velocity at the mass-shedding limit
is by about 20\% larger than that at the closest separation we can
calculate for each sequence where the less massive star has
$\chi \sim 0.6$. The orbital angular velocity at the mass-shedding
limit computed by the extrapolation is summarized in Appendix
\ref{app:omegamassshed}.

Let us continue investigating the orbital angular velocity at the
mass-shedding limit, $M_0 \Omega_{\rm ms}$. By regarding the Newtonian
quantities, $M_{\rm tot}$, $M_{\rm NS1}$, $M_{\rm NS2}$, and
$r_{\rm NS1}$, in Equation (\ref{eq:omega_tid}) as the relativistic ones,
$M_0$, $M_{\rm ADM}^{\rm NS1}$, $M_{\rm ADM}^{\rm NS2}$, and
$R_{\rm NS1}$, where $R_{\rm NS1}$ is the circumferential radius of
the less massive star, we can write $M_0 \Omega_{\rm ms}$ by an
empirical formula as
\begin{equation}
  M_0 \Omega_{\rm ms} = B~ {\cal C}_{\rm NS1}^{3/2}
  \Bigl( 1+\frac{1}{q} \Bigr)^{3/2} q^{1/2},
  \label{eq:omega_ms}
\end{equation}
where $q \equiv M_{\rm ADM}^{\rm NS1}/M_{\rm ADM}^{\rm NS2} \le 1$ is
the mass ratio,
${\cal C}_{\rm NS1} \equiv M_{\rm ADM}^{\rm NS1}/R_{\rm NS1}$ is the
compactness of the less massive neutron star, and 
$B \equiv (1/A)^{3/2}$ is a constant. Figure \ref{fig21} plots all
equal-mass data as a function of the compactness of a neutron star,  
and we perform the fitting procedure by assuming the form of
Equation (\ref{eq:omega_ms}). By this fitting, we determine the value of
constant as $B \simeq 0.260$. Note here that we do not find any
evidence that the value of $B$ depends on the EOS. In other words, 
the value of $M_0 \Omega_{\rm ms}$ depends only on the compactness of a
neutron star, and only weakly on the EOSs. 

\begin{figure}[t]
\epsscale{1.0}
\plotone{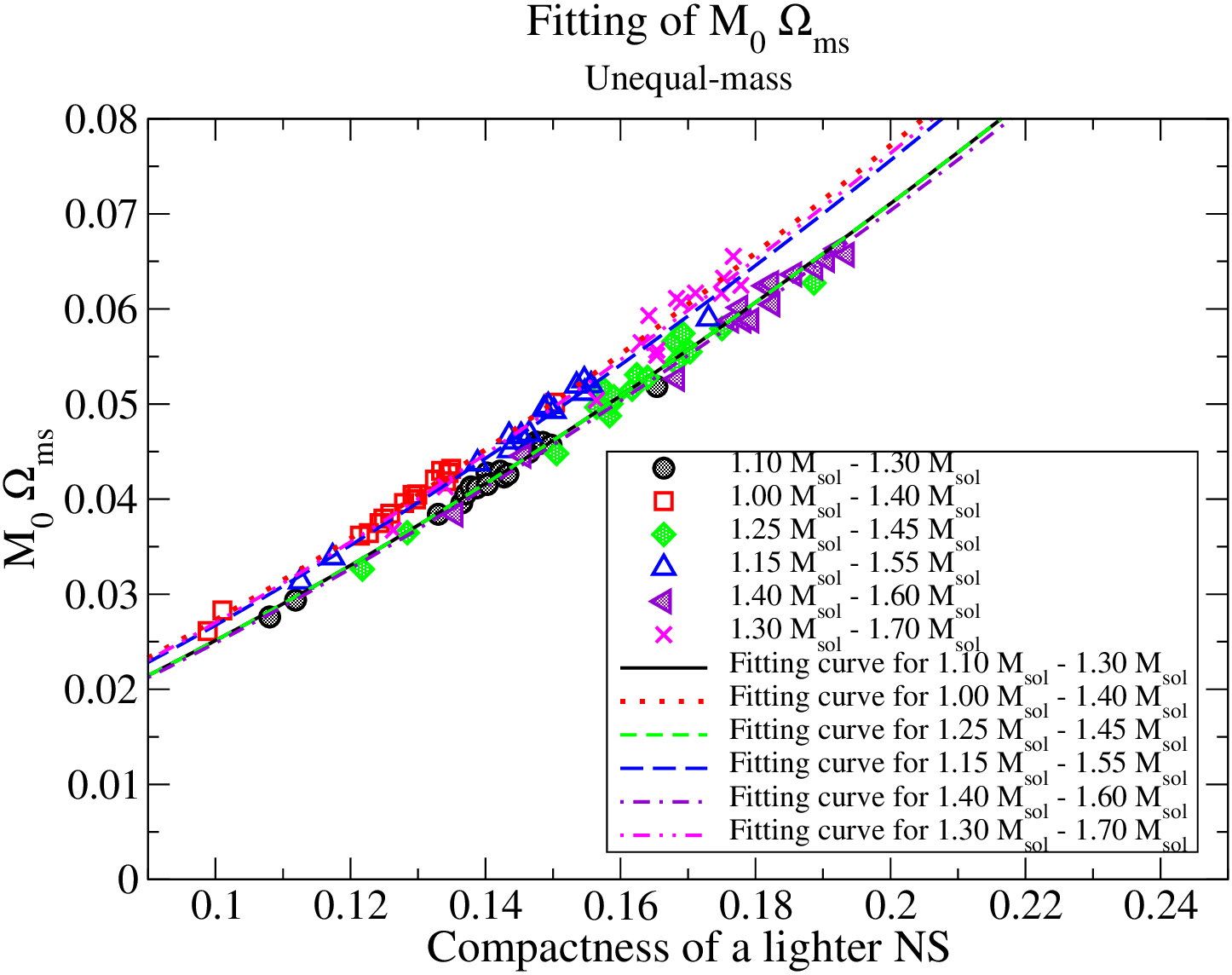}
\caption{Same as Figure \ref{fig21} but for unequal-mass models.
  Filled (black) circles, open (red) squares, filled (green) diamonds,
  open (blue) up-triangles, filled (violet) left-triangles, and
  (magenta) crosses denote, respectively, the orbital angular velocity
  at the mass-shedding limit for models of
  $1.10 M_{\odot}$ vs. $1.30 M_{\odot}$,
  $1.00 M_{\odot}$ vs. $1.40 M_{\odot}$,
  $1.25 M_{\odot}$ vs. $1.45 M_{\odot}$,
  $1.15 M_{\odot}$ vs. $1.55 M_{\odot}$,
  $1.40 M_{\odot}$ vs. $1.60 M_{\odot}$, and
  $1.30 M_{\odot}$ vs. $1.70 M_{\odot}$.
  The (black) solid, (red) dotted, (green) short-dashed, (blue)
  long-dashed, (violet) dot-dashed, and (magenta) dot-dot-dashed
  curves are the fitting curves for the data.
\label{fig22}}
\end{figure}

Figure \ref{fig22} shows the results of fitting of $M_0 \Omega_{\rm
ms}$ for unequal-mass binaries as a function of the compactness of a
less massive neutron star. For the fitting, we use the form of
Equation (\ref{eq:omega_ms}) and determine the constant $B$ for each mass
ratio. The obtained values are $B \simeq 0.268$ for the case of $1.10
M_{\odot}$ versus $1.30 M_{\odot}$ ($q \simeq 0.846$), $B \simeq 0.275$
for $1.00 M_{\odot}$ versus $1.40 M_{\odot}$ ($q \simeq 0.714$), $B \simeq
0.270$ for $1.25 M_{\odot}$ versus $1.45 M_{\odot}$ ($q \simeq 0.862$), $B
\simeq 0.273$ for $1.15 M_{\odot}$ versus $1.55 M_{\odot}$ ($q \simeq
0.742$), $B \simeq 0.268$ for $1.40 M_{\odot}$ versus $1.60 M_{\odot}$
($q=0.875$), and $B \simeq 0.279$ for $1.30 M_{\odot}$ versus $1.70
M_{\odot}$ ($q \simeq 0.765$). Even though the value $B$ depends
weakly on the mass ratio, i.e., the case of smaller $q$ tends to have
a larger $B$, we suppose that the dependence comes from the error of
extrapolation because the deviation of $B$ from its averaged value,
0.270, is about 3\%. We would like to remind that the estimated
orbital angular velocity at the mass-shedding limit is about 20\%
larger than that at the closest separation we can calculate: the error
of 3\% could be produced by a slight change in the extrapolation
curve. From the results for the mass ratio $0.71 < q \le 1$, we
conclude that the constant $B$ which appears in
Equation (\ref{eq:omega_ms}) can be set to $B=0.270$. 

The value of $B=0.270$ is slightly larger than that derived for
Newtonian close binaries \citep{pac71}. If we translate the constant
in \citet{pac71} for an equal-mass binary, it becomes $B=0.38^{3/2}
\simeq 0.23$; our result is about 17\% larger. On the other hand, the
value of $B=0.270$ we obtained is the same as that found for
quasi-equilibrium sequences of black hole-neutron star binaries in
general relativity in \citet{tan08}. In addition, \citet{shi06b,shi07}
gives the same value for a black hole-neutron star binary in general
relativity where the neutron star is corotating. The value of
$B=0.270$ with Equation (\ref{eq:omega_ms}) could be widely used as an
estimation of the orbital angular velocity at the mass-shedding limit
for neutron stars in a relativistic binary system. 

\begin{figure}[t]
\epsscale{1.0}
\plotone{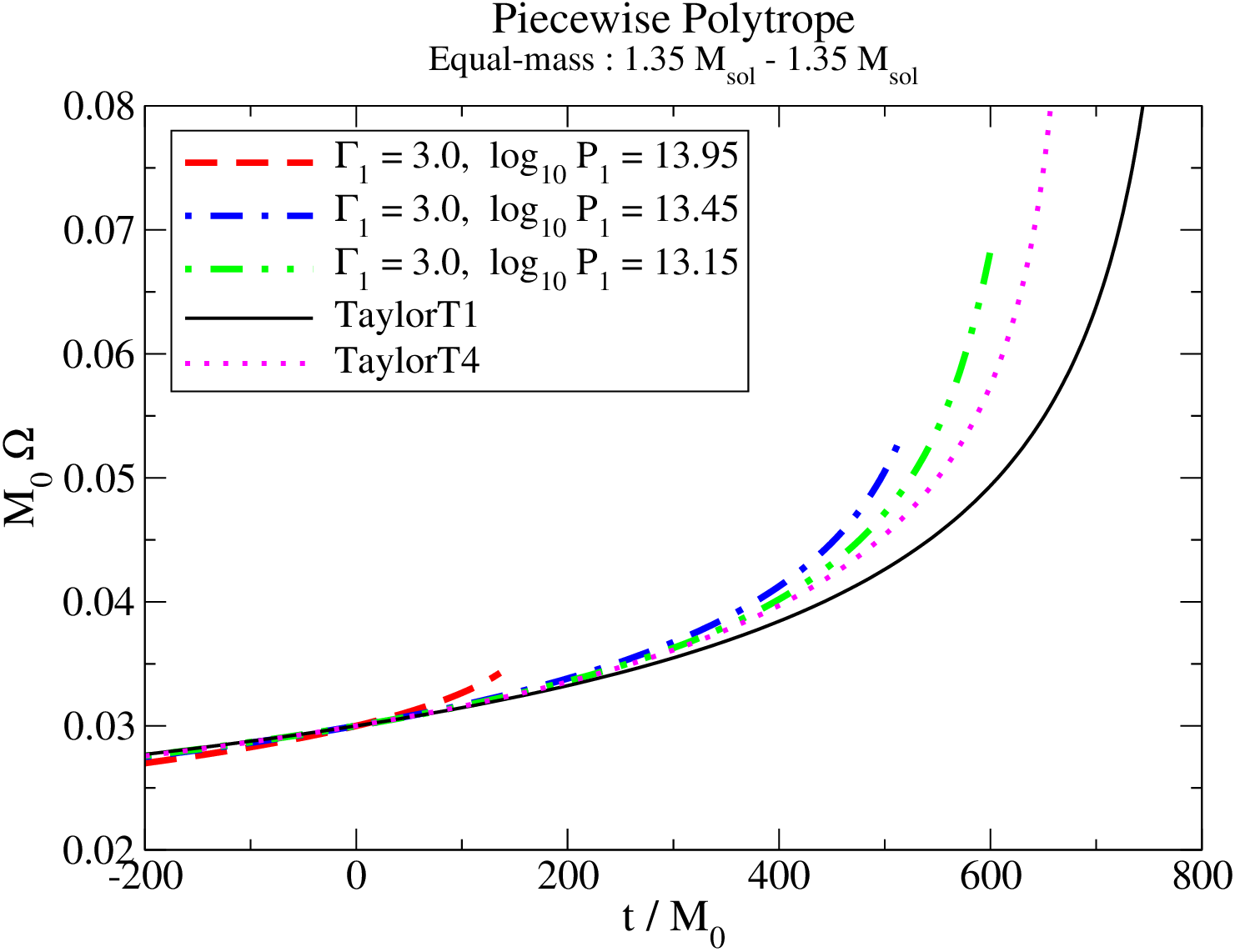}
\caption{Orbital angular velocity as a function of time for
  equal-mass binary neutron stars with a mass of $1.35 M_{\odot}$.
  All the EOSs are the piecewise polytrope with $\Gamma_1=3.0$
  but have different values of $\log_{10} P_1$. The thick (red)
  dashed, thick (blue) dot-dashed, and thick (green) dot-dot-dashed
  curves denote the cases of $\log_{10} P_1=13.95$ (PwPoly30-1395),
  13.45 (PwPoly30-1345), and 13.15 (PwPoly30-1315), respectively. The
  thin (black) solid and thin (magenta) dotted curves are calculated
  by using TaylorT1 and TaylorT4 in \citet{boy07}.
\label{fig23}}
\end{figure}

\subsection{Quasi-stationary evolution of the orbital angular velocity}

As we stated in Section 1, one of the primary purposes for constructing
quasi-equilibrium states is to provide initial data for inspiral and
merger simulations in numerical relativity. It will be quite useful
for numerical relativity community, if we provide quantities with
which they can compare the results of simulations in the inspiral
phase. One of the quantities derived from quasi-equilibrium results is
the time evolution of the orbital angular velocity. To obtain the
orbital angular velocity as a function of time, we adopt a similar
method introduced by \citet{boy07}. Assuming the quasi-stationary
adiabatic evolution of the inspiral orbit, we may write the time
derivative of the orbital angular velocity as
\begin{equation}
  \frac{d \Omega}{dt} =\frac{dE/dt}{dE/d\Omega}
  \equiv F(\Omega)^{-1}. \label{eq:dodt}
\end{equation}
This equation is the same as Equation (55) in \citet{ury09}. For the
temporal change in the binding energy of the binary neutron stars,
$dE/dt$, we use the 3.5PN equation shown in \citet{bla06}. On the
other hand, we estimate the change in the binding energy,
$dE/d\Omega$, as follows: we write the binding energy as
\begin{equation}
  E_{\rm b} = (E)_{\rm 3.5PN} + a x^5 + b x^6 + c x^7,
\end{equation}
where $x \equiv (M_0 \Omega)^{2/3}$, and $(E)_{\rm 3.5PN}$ denotes
the binding energy through 3.5PN order. The constants, $a$, $b$, and
$c$, are determined by numerically fitting the obtained sequences. Then we
calculate the change in the binding energy by taking the derivative
with respect to the orbital angular velocity as
\begin{equation}
  \frac{dE}{d\Omega} = \frac{d}{d\Omega} E_{\rm b}.
\end{equation}

\begin{figure}[t]
\epsscale{1.0}
\plotone{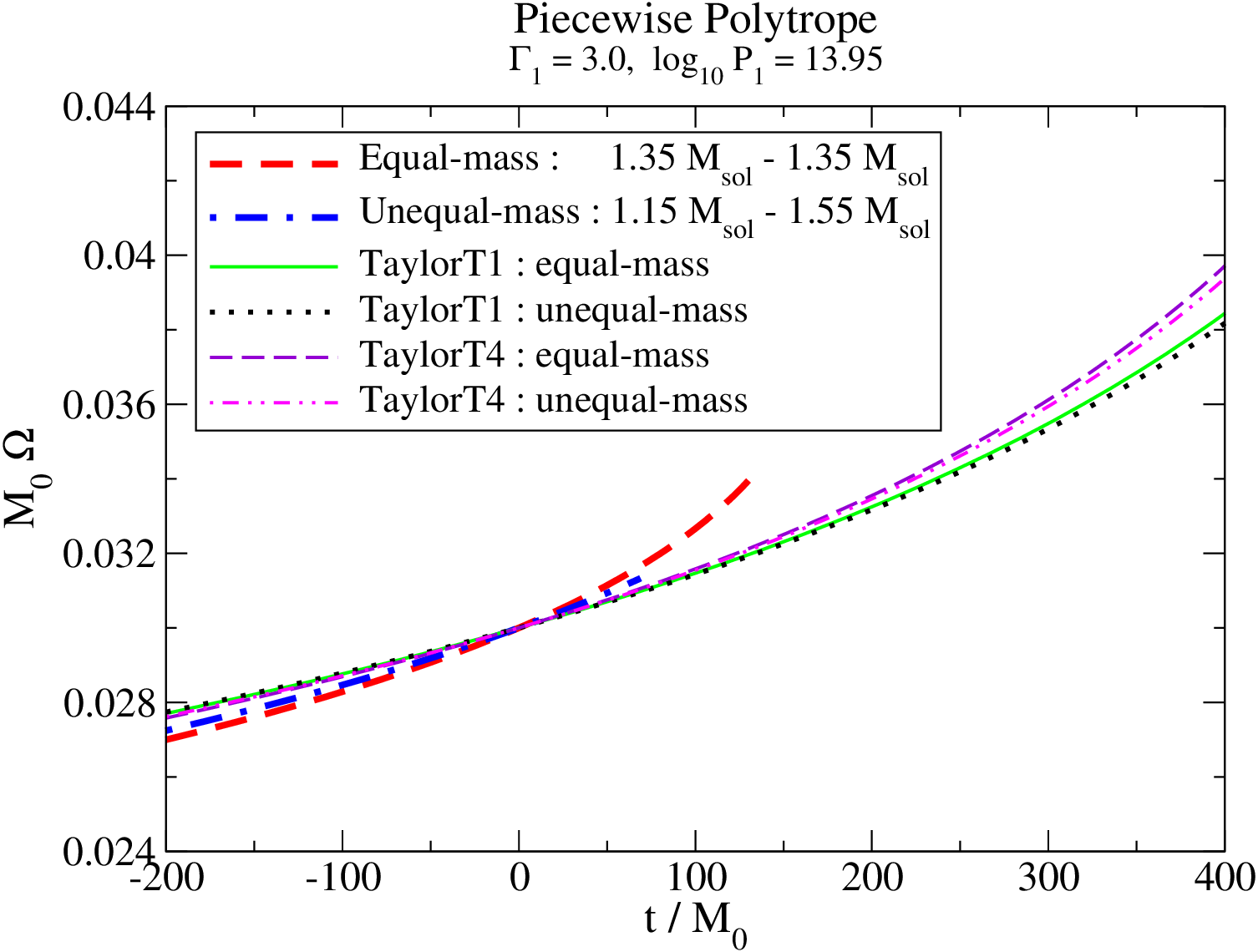}
\caption{Same as Figure \ref{fig23} but for the piecewise polytrope
  with $\Gamma_1=3.0$ and $\log_{10} P_1=13.95$ (PwPoly30-1395) and
  for comparing the results of equal-mass with unequal-mass binaries.
  The thick (red) dashed and thick (blue) dot-dashed curves denote the
  cases of equal-mass binary with a mass of $1.35 M_{\odot}$ and those
  of unequal-mass binary with masses of $1.15 M_{\odot}$ and
  $1.55 M_{\odot}$. The thin (green) solid and thin (black) dotted
  curves are calculated by using TaylorT1 in \citet{boy07} for
  equal-mass binary and unequal-mass one of mass ratio,
  $1.15/1.55$. The thin (violet) dashed and thin (magenta)
  dot-dot-dashed curves are for TaylorT4 in \citet{boy07}.
\label{fig24}}
\end{figure}

\begin{figure}[t]
\epsscale{1.0}
\plotone{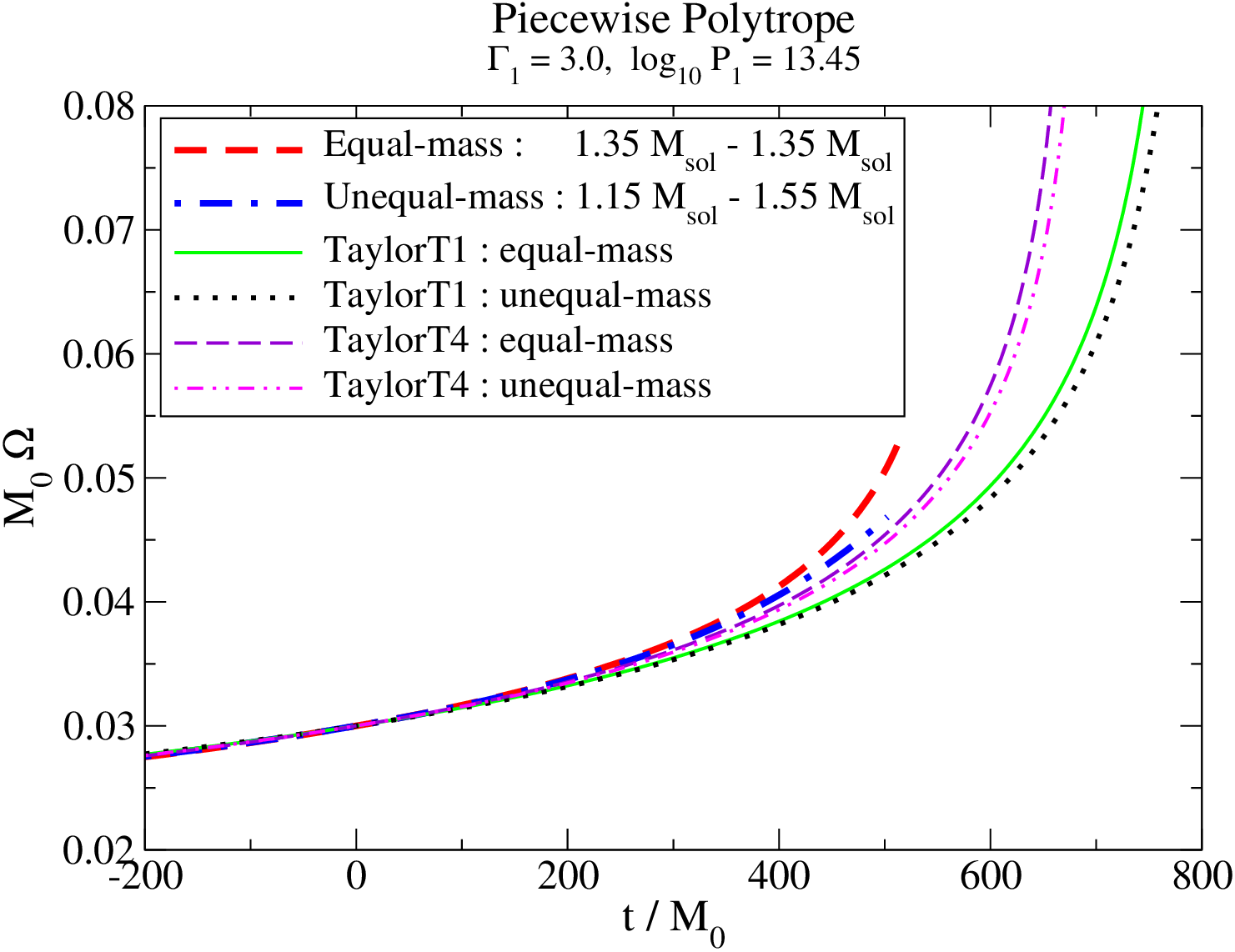}
\caption{Same as Figure \ref{fig24} but for $\Gamma_1=3.0$ and
  $\log_{10} P_1=13.45$ (PwPoly30-1345).
\label{fig25}}
\end{figure}

By numerically integrating Equation (\ref{eq:dodt}), we obtain the orbital
angular velocity as a function of time,
\begin{equation}
  t =\int_{\Omega_{\rm ini}}^{\Omega_{\rm fin}} F (\Omega)
  d \Omega.
\end{equation}
Here $\Omega_{\rm ini}$ is the initial orbital angular velocity at
which the time is defined to be zero, and $\Omega_{\rm fin}$ is the
final orbital angular velocity. For $\Omega_{\rm fin}$, we use the
orbital angular velocity at the mass-shedding limit obtained in
Section \ref{sec:endpoint}.

Figure \ref{fig23} shows the orbital angular velocity as a function of
time for three piecewise polytropes with $\Gamma_1=3.0$. The thick
(red) dashed, thick (blue) dot-dashed, and thick (green)
dot-dot-dashed curves denote the results for $\log_{10} P_1=13.95$
(PwPoly30-1395), 13.45 (PwPoly30-1345), and 13.15 (PwPoly30-1315),
respectively. All of them are calculated by using the equal-mass sequences
of $M_{\rm ADM}^{\rm NS1}=M_{\rm ADM}^{\rm NS2}=1.35 M_{\odot}$. We
also show two reference curves of TaylorT1 and TaylorT4 introduced by
\citet{boy07}. The initial orbital angular velocity is set to
$M_0 \Omega_{\rm ini}=0.03$. It is found that the curve of the most
compact neutron stars (the dot-dot-dashed curve, PwPoly30-1315) is the
closest to that of TaylorT4 among all the models. This is reasonable
because the neutron stars in this EOS are the most compact and tidal
effects are the weakest. On the other hand, the curve of the least
compact neutron stars (the dashed curve, PwPoly30-1395) has already
deviated from the curves of TaylorT1 and TaylorT4 before reaching the
initial orbital angular velocity $M_0 \Omega_{\rm ini}=0.03$, because of
the tidal deformation. This shows that if we would like to compare the
time evolution of the orbital angular velocity obtained by
simulations, we need to start the simulations from a much smaller
value of the orbital angular velocity for such less-compact neutron
star models.

\begin{figure}[t]
\epsscale{1.0}
\plotone{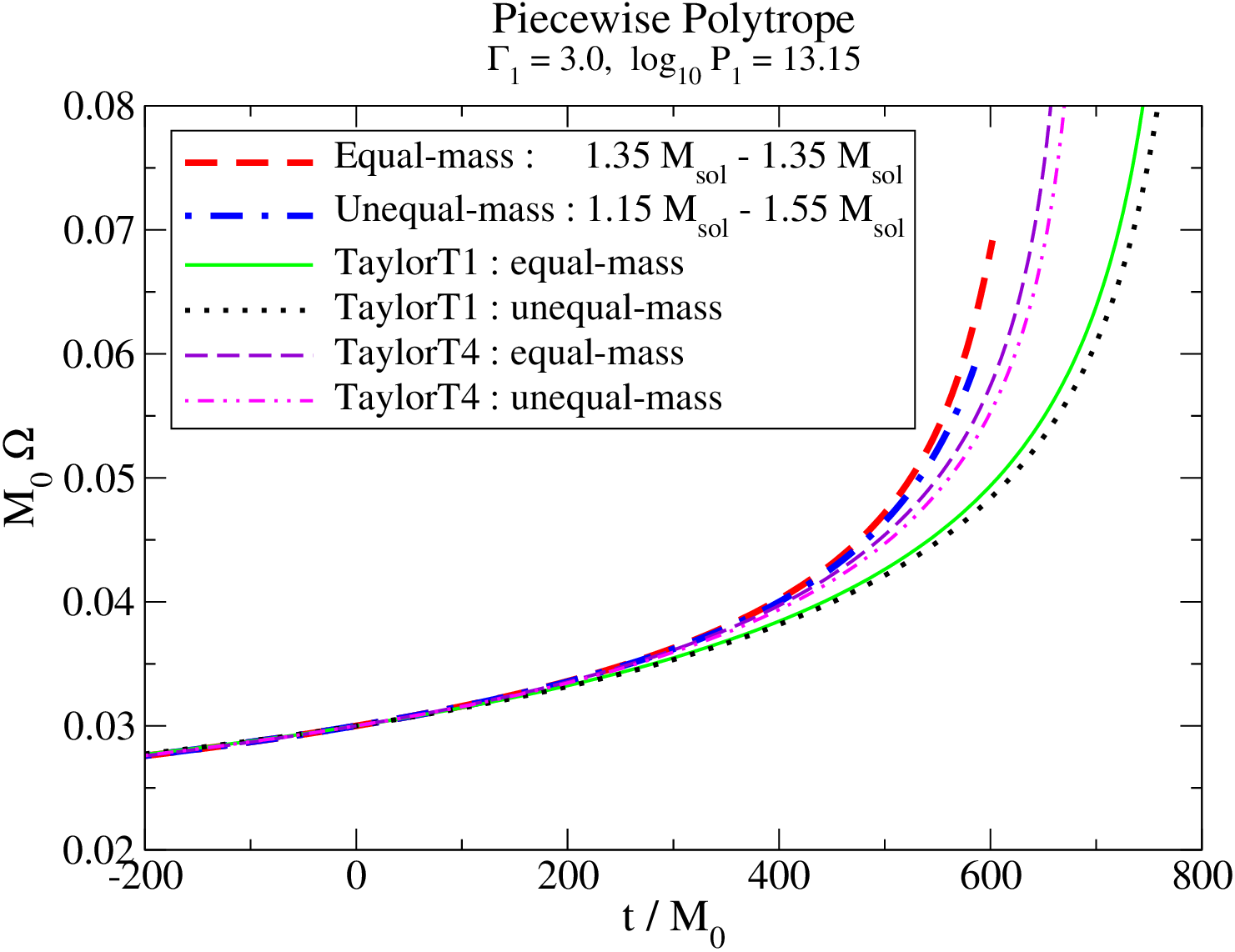}
\caption{Same as Figure \ref{fig24} but for $\Gamma_1=3.0$ and
  $\log_{10} P_1=13.15$ (PwPoly1315).
\label{fig26}}
\end{figure}

Figure \ref{fig23} also shows that the quasi-equilibrium results are
closer to those of TaylorT4 than those of TaylorT1, even though our
calculation method of the orbital angular velocity as a function of
time is similar to that of TaylorT1 (see Equation (\ref{eq:dodt})). We,
however, believe that at the limit of test mass, our method gives
similar results of TaylorT1.

Figures \ref{fig24}--\ref{fig26} compare the results of unequal-mass
binaries with those of equal-mass ones with the total mass of $M_0=2.7
M_{\odot}$. The equal-mass binary has the mass of $M_{\rm ADM}^{\rm
NS1}=M_{\rm ADM}^{\rm NS2}=1.35 M_{\odot}$, and the unequal-mass
binary does $M_{\rm ADM}^{\rm NS1}=1.15 M_{\odot}$ and $M_{\rm
ADM}^{\rm NS2}=1.55 M_{\odot}$.  The EOSs are the same as those in
Figure \ref{fig23}, but the results for $\log_{10} P_1=13.95$
(PwPoly30-1395), 13.45 (PwPoly30-1345), and 13.15 (PwPoly30-1315) are
shown in Figures \ref{fig24}--\ref{fig26}, respectively. From these
figures, we find that the curves for unequal-mass binaries are located
below those for equal-mass binaries for $t > 0$ and are closer to the
post-Newtonian results, in particular for a less-compact neutron star
model (see Figures \ref{fig25} and \ref{fig26}). It is also found that
the curves for unequal-mass binaries terminate before their deviation
from the post-Newtonian results becomes larger. These two findings
imply that even though the deformation of the less massive star is
larger in the case of unequal-mass binaries than that in the
equal-mass case when we compare at the same orbital angular velocity,
the more massive star remains close to a spherical star and dominates
the motion of binary system in the unequal-mass case. The effect of
the deformation of stars on the orbital angular velocity is larger for
the equal-mass binaries if the total mass is the same.

\subsection{Comparison of the results by tabulated realistic
EOSs with those by a fitting formula}

Before closing Section \ref{sec:results}, we would like to comment on the
results by using the fitted EOSs. As shown in Figure \ref{fig3}, the
mass-radius relation is very close to that for the tabulated realistic
EOSs because we mimic them in constructing the fitting formula
\citep{shi05,shi06a}. As expected, the difference between the results
for the fitted EOSs and those for the tabulated realistic EOSs is very
small. Even for the orbital angular velocity at the mass-shedding
limit which is obtained by the extrapolation of sequences, the
difference between the results of these two EOSs is less than about
5\%.

\section{Summary}

In the present paper, quasi-equilibrium sequences of binary neutron
stars are constructed for 18 EOSs except for the $\Gamma=2$ polytrope,
in the IWM framework in general relativity. The EOSs we choose are
nine piecewise polytropes, six tabulated realistic EOSs derived by
using various theories of dense nuclear matter and different solution
methods of the many-body problem in nuclear physics, and three EOSs
expressed by a fitting formula. We employ three total masses,
$M_0=2.4 M_{\odot}$, $2.7 M_{\odot}$, and $3.0 M_{\odot}$ for each
EOS, and compute sequences of three mass ratios for each total mass:
i.e., $M_{\rm ADM}^{\rm NS1}$ versus $M_{\rm ADM}^{\rm NS2}$
$=1.20 M_{\odot}$ versus $1.20 M_{\odot}$,
$1.10 M_{\odot}$ versus $1.30 M_{\odot}$, and
$1.00 M_{\odot}$ versus $1.40 M_{\odot}$ for $M_0=2.4 M_{\odot}$;
$1.35 M_{\odot}$ versus $1.35 M_{\odot}$,
$1.25 M_{\odot}$ versus $1.45 M_{\odot}$, and
$1.15 M_{\odot}$ versus $1.55 M_{\odot}$ for $M_0=2.7 M_{\odot}$;
and $1.50 M_{\odot}$ versus $1.50 M_{\odot}$,
$1.40 M_{\odot}$ versus $1.60 M_{\odot}$, and
$1.30 M_{\odot}$ versus $1.70 M_{\odot}$ for $M_0=3.0 M_{\odot}$.

We focus on the unequal-mass sequences and compare their results with
those of the equal-mass case. Changing the mass ratio, the total mass,
and the EOSs, we investigate the behavior of the binding energy and
total angular momentum along a sequence, the endpoint of sequences,
and the orbital angular velocity as a function of time. For example,
it is found for the piecewise polytropic EOSs that the change in
stellar radius fixing the stiffness of the core EOS makes the orbital
angular velocity at the mass-shedding limit vary widely, while the
change in stiffness of the core EOS fixing the stellar radius does not
change the orbital angular velocity at the mass-shedding limit
significantly.

It is also found that the orbital angular velocity at the closest
separation decreases as we decrease the mass ratio,
$M_{\rm ADM}^{\rm NS1}/M_{\rm ADM}^{\rm NS2} \le 1$. The reason is
that the less massive star in an unequal-mass binary is tidally
deformed by the companion more massive star and starts shedding mass
at larger separation than that for the equal-mass case. It is
found that the orbital angular velocity at the mass-shedding limit
increases as the neutron-star mass increases. This is because a more
massive star becomes more compact and more difficult to be tidally
disrupted for the same EOS. This implies that the binary neutron
stars with massive stars need to come closer than those with less
massive stars for reaching the mass-shedding limit. The orbital
angular velocity at the mass-shedding limit is analyzed by using a
Newtonian argument, and an empirical formula is found as
\begin{equation}
  M_0 \Omega_{\rm ms} =0.270~ {\cal C}_{\rm NS1}^{3/2}
  \Bigl( 1+\frac{1}{q} \Bigr)^{3/2} q^{1/2}.
\end{equation}

We have provided tables for 160 sequences as shown in Appendix
\ref{app:sequence}. Those tables may be useful as one of the database
for future works on binary neutron stars in quasi-equilibrium and as a
guideline of numerical simulations for the inspiral and merger.

\acknowledgments

We thank Charalampos Markakis and Koutarou Kyutoku for providing a
list of parameters for the piecewise polytropic equations of state.
We also thank John L. Friedman for useful comments. This work was
supported in part by NSF Grant PHY-0503366, and by Grant-in-Aid for
Scientific Research (21340051) and for Scientific Research on
Innovative Area (20105004) of the Japanese MEXT.

\appendix

\section{Data of spherical stars} \label{app:spherical}

In Table \ref{table2}, some selected data of
spherical neutron stars is listed. In each table, the ADM mass
$M_{\rm ADM}^{\rm NS}$, the baryon rest mass $M_{\rm B}$, the
circumferential radius $R_{\rm NS}$, the coordinate radius
$a_{\rm NS}$, the compactness
${\cal C} \equiv M_{\rm ADM}^{\rm NS}/R_{\rm NS}$, the central baryon
rest-mass density $\rho_c$ in cgs units, and the central specific
internal energy $\epsilon_c$ are shown. The number in parentheses for
$\rho_c$ means the exponent of 10. For each EOS, we show
the data at the maximum mass in the last line.

\section{Sequence data} \label{app:sequence}

The sequence data for the polytropic EOS with $\Gamma=2$ is summarized
in Table \ref{table3}, that for the piecewise polytropic EOSs,
the tabulated realistic EOSs, and the EOSs with a fitting formula
in Table \ref{table4}, and that for the piecewise polytropic EOSs
calculated by the old code in Table \ref{table5}.
The coordinate separation $d/M_0$, the
orbital angular velocity $M_0 \Omega$, the binding energy
$E_{\rm b}/M_0$, the total angular momentum $J/M_0^2$, the
mass-shedding indicator for the less massive star $\chi_1$, that for
the more massive star $\chi_2$, and the virial error $\delta M$ are
shown. For the sequences for the $\Gamma=2$ polytrope, we additionally
show the coordinate separation in polytropic units $\bar{d}$. For the
sequences of equal-mass binaries, we only show the mass-shedding
indicator for one neutron star $\chi_1$ because the two stars are
identical. However, because of numerical error there is a slight
difference between the values of $\chi$ for two stars. Such a
difference is described by
$\delta \chi \equiv |(\chi_1 -\chi_2)/\chi_1|$ in the tables for
equal-mass binaries. The number in parentheses means the exponent of
10.

\section{Orbital angular velocity at the mass-shedding limit}
\label{app:omegamassshed}

In Table \ref{table6}, we summarize the orbital angular velocity at
the mass-shedding limit which is estimated by using the extrapolation
method described in Section \ref{sec:endpoint}. The orbital angular
velocity is listed for each EOS and for each mass ratio. There
are three cases for which we do not show the estimated value, i.e., for the
cases of PwPoly27-1335, FPS, and fitFPS with masses of
$1.50 M_{\odot}$ versus $1.50 M_{\odot}$. For those cases, we cannot
calculate configurations for close separations, because both stars
are compact and it is difficult to converge the computation. Since the
mass-shedding indicator $\chi$ is still larger than 0.8 for those
sequences at the closest separation we can calculate, the
extrapolation method does not work well for estimating the orbital
angular velocity at the mass-shedding limit.



\begin{deluxetable}{cccccccc}
\tabletypesize{\scriptsize}
\tablecaption{Selected data of spherical stars with piecewise
  polytropic EOSs, tabulated realistic EOSs, and fitted EOSs
  \label{table2}}
\tablewidth{0pt}
\tablehead{
\colhead{EOS name} &
\colhead{$M_{\rm ADM}^{\rm NS}$ [$M_{\odot}$]} &
\colhead{$M_{\rm B}$ [$M_{\odot}$]} &
\colhead{$R_{\rm NS}$ [km]} & \colhead{$a_{\rm NS}$ [km]} &
\colhead{${\cal C}$} & \colhead{$\rho_c$ [${\rm g~cm^{-3}}$]} &
\colhead{$\epsilon_c$}
}
\startdata
PwPoly30-1395 &
  1.150 & 1.225 & 15.08 & 13.33 & 0.1126 & 3.378(+14) & 1.048 \\
& 1.250 & 1.339 & 15.16 & 13.25 & 0.1218 & 3.560(+14) & 1.053 \\
& 1.350 & 1.455 & 15.23 & 13.16 & 0.1309 & 3.743(+14) & 1.058 \\
& 1.450 & 1.572 & 15.28 & 13.05 & 0.1401 & 3.930(+14) & 1.063 \\
& 1.550 & 1.691 & 15.32 & 12.93 & 0.1494 & 4.124(+14) & 1.068 \\
& 2.835 & 3.409 & 13.37 & 8.677 & 0.3131 & 1.085(+15) & 1.425 \\
\tableline
PwPoly30-1355 &
  1.150 & 1.245 & 12.23 & 10.47 & 0.1388 & 6.234(+14) & 1.065 \\
& 1.250 & 1.363 & 12.26 & 10.33 & 0.1505 & 6.623(+14) & 1.072 \\
\enddata
\end{deluxetable}

\begin{deluxetable}{cccccccc}
\tabletypesize{\small}
\tablecaption{Quasi-equilibrium sequence data for the polytropic EOS
  with $\Gamma=2$ \label{table3}}
\tablewidth{0pt}
\tablehead{
\colhead{$\bar{d}$} & \colhead{$d/M_0$} & \colhead{$M_0 \Omega$} &
\colhead{$E_{\rm b}/M_0$} & \colhead{$J/M_0^2$} &
\colhead{$\chi_1$} & \colhead{$\delta \chi$ or $\chi_2$} &
\colhead{$\delta M$}
}
\startdata
\cutinhead{$\bar{M}_{\rm B}=$ 0.15 vs. 0.15}
4.872 & 17.46 & 1.270(-2) & -6.481(-3) & 1.1675 &
0.9755 & 8.375(-10) & 1.285(-5) \\
4.263 & 15.28 & 1.535(-2) & -7.294(-3) & 1.1094 &
0.9612 & 1.812(-9)~ & 1.403(-5) \\
3.959 & 14.19 & 1.705(-2) & -7.779(-3) & 1.0795 &
0.9499 & 3.374(-9)~ & 1.463(-5) \\
3.654 & 13.10 & 1.909(-2) & -8.331(-3) & 1.0490 &
0.9336 & 7.775(-9)~ & 1.572(-5) \\
3.350 & 12.01 & 2.156(-2) & -8.964(-3) & 1.0179 &
0.9093 & 2.219(-8)~ & 1.894(-5) \\
3.045 & 10.91 & 2.463(-2) & -9.692(-3) & 0.9865 &
0.8705 & 1.182(-8)~ & 2.438(-5) \\
2.862 & 10.26 & 2.684(-2) & -1.018(-2) & 0.9676 &
0.8347 & 4.197(-8)~ & 3.673(-5) \\
2.680 & 9.605 & 2.939(-2) & -1.073(-2) & 0.9487 &
0.7816 & 5.520(-9)~ & 8.642(-5) \\
\enddata
\end{deluxetable}

\begin{deluxetable}{cccccccccc}
\tabletypesize{\footnotesize}
\tablecaption{Quasi-equilibrium sequence data for the piecewise
polytropic EOSs, the tabulated realistic EOSs, and the EOSs with
a fitting formula \label{table4}}
\tablewidth{0pt}
\tablehead{
\colhead{EOS name} & \colhead{$M_0$} &
\colhead{$M_{\rm ADM}^{\rm NS 1}$ vs. $M_{\rm ADM}^{\rm NS 2}$} &
\colhead{$d/M_0$} & \colhead{$M_0 \Omega$} &
\colhead{$E_{\rm b}/M_0$} & \colhead{$J/M_0^2$} &
\colhead{$\chi_1$} & \colhead{$\delta \chi$ or $\chi_2$} &
\colhead{$\delta M$}
}
\startdata
\cutinhead{Piecewise polytropic EOSs}
PwPoly30-1395 & $2.4 M_{\odot}$ & $1.20 M_{\odot}$ vs. $1.20 M_{\odot}$
& 22.57 & 8.788(-3) & -5.137(-3) & 1.2952 & 0.9763 & 7.598(-11) &
1.528(-6) \\
&&& 19.75 & 1.065(-2) & -5.801(-3) & 1.2267 & 0.9626 & 6.139(-11) &
3.168(-6) \\
&&& 18.34 & 1.185(-2) & -6.200(-3) & 1.1912 & 0.9517 & 1.166(-10) &
4.497(-6) \\
&&& 16.93 & 1.328(-2) & -6.657(-3) & 1.1549 & 0.9359 & 5.224(-11) &
6.882(-6) \\
&&& 15.52 & 1.504(-2) & -7.183(-3) & 1.1178 & 0.9122 & 1.532(-10) &
1.252(-5) \\
&&& 14.11 & 1.722(-2) & -7.792(-3) & 1.0801 & 0.8738 & 1.578(-10) &
2.245(-5) \\
&&& 13.26 & 1.880(-2) & -8.208(-3) & 1.0574 & 0.8376 & 1.038(-10) &
4.047(-5) \\
&&& 12.70 & 1.999(-2) & -8.508(-3) & 1.0423 & 0.8039 & 1.902(-12) &
6.155(-5) \\
\enddata
\end{deluxetable}

\begin{deluxetable}{cccccccc}
\tabletypesize{\footnotesize}
\tablecaption{Quasi-equilibrium sequence data for the piecewise
  polytropic EOSs calculated by the old code \label{table5}}
\tablewidth{0pt}
\tablehead{
\colhead{EOS name} &
\colhead{$d/M_0$} & \colhead{$M_0 \Omega$} &
\colhead{$E_{\rm b}/M_0$} & \colhead{$J/M_0^2$} &
\colhead{$\chi_1$} & \colhead{$\delta \chi$} &
\colhead{$\delta M$}
}
\startdata
\cutinhead{$M_0 =2.7 M_{\odot}$ ($1.35 M_{\odot}$ vs. $1.35 M_{\odot}$)}
PwPoly30-1395 &
20.06 & 1.042(-2) & -5.716(-3) & 1.2349 & 0.9777 & 7.312(-11) &
2.312(-5) \\
& 17.56 & 1.261(-2) & -6.445(-3) & 1.1712 & 0.9647 & 6.789(-13) &
2.364(-5) \\
& 16.30 & 1.402(-2) & -6.883(-3) & 1.1382 & 0.9543 & 4.033(-11) &
2.397(-5) \\
& 15.05 & 1.571(-2) & -7.382(-3) & 1.1046 & 0.9393 & 1.640(-12) &
2.436(-5) \\
& 13.79 & 1.777(-2) & -7.954(-3) & 1.0703 & 0.9168 & 7.662(-12) &
2.480(-5) \\
& 12.54 & 2.032(-2) & -8.615(-3) & 1.0355 & 0.8807 & 2.144(-12) &
2.527(-5) \\
& 11.79 & 2.218(-2) & -9.057(-3) & 1.0146 & 0.8471 & 2.353(-11) &
2.553(-5) \\
& 11.29 & 2.358(-2) & -9.372(-3) & 1.0008 & 0.8161 & 2.248(-10) &
2.565(-5) \\
\enddata
\end{deluxetable}

\begin{deluxetable}{ccccc}
\tabletypesize{\footnotesize}
\tablecaption{Orbital angular velocity at the mass-shedding limit
  \label{table6}}
\tablewidth{0pt}
\tablehead{
\colhead{EOS name} &
\colhead{$M_0 [M_{\odot}]$} &
\colhead{$M_{\rm ADM}^{\rm NS1} [M_{\odot}]$} &
\colhead{$M_{\rm ADM}^{\rm NS2} [M_{\odot}]$} &
\colhead{$M_0 \Omega_{\rm ms}$}
}
\startdata
\cutinhead{Piecewise Polytropic EOSs}
PwPoly30-1395
& 2.40 & 1.20 & 1.20 & 0.0289 \\
&      & 1.10 & 1.30 & 0.0276 \\
&      & 1.00 & 1.40 & 0.0261 \\
& 2.70 & 1.35 & 1.35 & 0.0343 \\
&      & 1.25 & 1.45 & 0.0326 \\
&      & 1.15 & 1.55 & 0.0313 \\
& 3.00 & 1.50 & 1.50 & 0.0402 \\
&      & 1.40 & 1.60 & 0.0384 \\
\enddata
\end{deluxetable}

\end{document}